\documentclass[aps, pre, 10pt]{revtex4-2}
\usepackage{etoolbox}
\usepackage{graphics,graphicx} 
\usepackage{natbib}
\usepackage{subfig}
\usepackage{float}
\usepackage{placeins}
\usepackage{ragged2e}
\usepackage{silence}
\usepackage{caption}
\usepackage{xcolor}
\usepackage{amsmath}
\usepackage{amssymb}
\usepackage{tikz}
\usepackage{braket}
\usepackage{blindtext}
\usepackage{bm}
\usepackage{subfig}
\usepackage{algorithm}
\usepackage{algpseudocode}
\usepackage{hyperref}
\hypersetup{
    colorlinks=true,
    citecolor=red,
    linkcolor=blue,
    urlcolor=black
}
\usepackage{MnSymbol}
\usepackage[cal=boondox,scr=boondoxo]{mathalfa}
\usepackage{booktabs}
\begin{document}


\title{Classical observable coalescence as a discriminating diagnostic for spectral degeneracies in non-Hermitian systems}


\author{P.K. Nkaimbi}
\affiliation{Department of Physics, Higher Teachers Training College, University of Bamenda, P.O.BOX: 11 Bamenda, Cameroon}

\affiliation{Promotion Centre of Research for Technological Advancement and Sustainable Development (PCR-TASD), P.O.Box: 55 Maroua-Cameroon}
\affiliation{Quantum Materials and Computing Group (QMaCG), P.O.Box: 70 Bambili, North West Region, Cameroon}

\author{C.A. Mba}
\affiliation{Laboratoire d'Optique, Laser et Applications; Universit\'e des Sciences et Techniques de Masuku, B. P. 943, Franceville, Gabon.}
\affiliation{ D\'epartement de Physique; Facult\'e des Sciences; Universit\'e des Sciences et Techniques de Masuku, B. P. 943, Franceville, Gabon.}

\author{A. B. Moubissi}
\affiliation{Laboratoire d'Optique, Laser et Applications; Universit\'e des Sciences et Techniques de Masuku, B. P. 943, Franceville, Gabon.}
\affiliation{ D\'epartement de Physique; Facult\'e des Sciences; Universit\'e des Sciences et Techniques de Masuku, B. P. 943, Franceville, Gabon.}

\author{M.R. Kamsap}
\affiliation{Laboratoire d'Optique, Laser et Applications, Unité de Recherche en Physique, Université des Sciences et Techniques de Masuku, B.P. 943, Franceville, Gabon}
\affiliation{Département de Physique, Faculté des Sciences, Université des Sciences et Techniques de Masuku, B.P. 943, Franceville, Gabon}
\author{L.C. Fai}
\affiliation{Unité de Recherche de Matière Condensée, d'\'Electronique et de Traitement de Signal (URMACETS), Department of Physics, Faculty of Science, University of Dschang, Cameroon, P.O.Box: 67 Dschang-Cameroon}
\affiliation{Quantum Materials and Computing Group (QMaCG), P.O.Box: 70 Bambili, North West Region, Cameroon}

\author{M.E. Ateuafack}
\affiliation{Department of Electrical and Electronics Engineering, College Of Technology, University of Buea, P.O. BOX: 63 Buea, Cameroon}
\affiliation{Unité de Recherche de Matière Condensée, d'\'Electronique et de Traitement de Signal (URMACETS), Department of Physics, Faculty of Science, University of Dschang, Cameroon, P.O.Box: 67 Dschang-Cameroon}
\affiliation{Promotion Centre of Research for Technological Advancement and Sustainable Development (PCR-TASD), P.O.Box: 55 Maroua-Cameroon}
\affiliation{Quantum Materials and Computing Group (QMaCG), P.O.Box: 70 Bambili, North West Region, Cameroon}

\author{M.N. Jipdi}
\email[]{jmichaelnicky@yahoo.fr}
\affiliation{Department of Physics, Higher Teachers Training College, University of Bamenda, P.O.BOX: 11 Bamenda, Cameroon}
\affiliation{Unité de Recherche de Matière Condensée, d'\'Electronique et de Traitement de Signal (URMACETS), Department of Physics, Faculty of Science, University of Dschang, Cameroon, P.O.Box: 67 Dschang-Cameroon}
\affiliation{Promotion Centre of Research for Technological Advancement and Sustainable Development (PCR-TASD), P.O.Box: 55 Maroua-Cameroon}
\affiliation{Quantum Materials and Computing Group (QMaCG), P.O.Box: 70 Bambili, North West Region, Cameroon}


\date{\today}

\begin{abstract}
Exceptional points (EPs) are conventionally identified through eigenvector
coalescence, inaccessible in classical platforms lacking a Hamiltonian
spectrum. We introduce classical observable coalescence as a diagnostic
identifying EPs directly from physical configurations, and demonstrate it
in a known non-Hermitian Ising chain with coupling $J$ and staggered
imaginary field $\gamma$, chosen for verifiability against established
quantum results. A mean-field reduction of the Heisenberg dynamics yields
equilibrium branches organising into two symmetry-related families whose
fusion at the boundary of the EP supersurface $|\gamma|=|J|$ exhibits the
characteristic $\sqrt{J^2-\gamma^2}$ branch-point structure and two-sheeted
Riemann topology. Crucially, the degeneracy at $\gamma=0,\eta =0 $ is shown to be a
diabolic point, producing no observable coalescence, establishing the
diagnostic as discriminating rather than merely suggestive. An exact
Jordan-Wigner and Bogoliubov-de Gennes treatment confirms eigenvector
coalescence at the same threshold, and a coupled gain-loss circuit
realisation is proposed, establishing classical observables as
experimentally accessible signatures of non-Hermitian criticality.

\end{abstract}


\maketitle


\section{Introduction}
\label{sec:introduction}

Non-Hermitian physics has emerged as a central framework for describing open,
dissipative, and driven systems, in which gain, loss, and asymmetric coupling
are not perturbations but defining features of the dynamics
\cite{bender1998real,el2007theory,ashida2020non,wah2025manybody,wah2026bridging}. Unlike Hermitian operators,
whose eigenvalues are real and eigenvectors orthogonal, non-Hermitian
operators admit complex spectra and non-orthogonal eigenstates, enabling
phenomena with no Hermitian counterpart. Among these, the exceptional point
(EP) occupies a privileged position: at an EP, two or more eigenvalues and
their corresponding eigenvectors coalesce simultaneously, rendering the
operator defective and its eigenbasis incomplete
\cite{heiss2012physics,moiseyev2011non,kato1966perturbation,rotter2009non}.
Near a second-order EP, the eigenvalues exhibit the characteristic Puiseux
expansion $\lambda_{\pm}(\mu)=\lambda_{\mathrm{EP}}\pm
C(\mu-\mu_{\mathrm{EP}})^{1/2}$, so that the spectral branches reside on
different sheets of a Riemann surface and are exchanged upon encircling the
branch point \cite{kato1966perturbation,berry2004physics,miri2019exceptional}.
This branch-point topology, together with the simultaneous coalescence of
eigenvalues and eigenvectors, distinguishes an EP from an ordinary
degeneracy, such as the diabolic point (DP) of a Hermitian conical
intersection, at which eigenvalues may coincide while the eigenvectors remain
linearly independent and orthogonally chosen
\cite{berry2004physics,heiss2012physics}.

The experimental relevance of EPs has been established across an
extraordinary range of platforms, including coupled optical resonators and
waveguides \cite{peng2014parity,feng2013experimental,ozdemir2019parity},
microwave cavities \cite{schindler2011experimental}, mechanical and acoustic
metamaterials \cite{coulais2017static,ryu2024exceptional}, and, more
recently, electronic circuits \cite{fernandez2026parity}. This universality
has motivated growing interest in classical non-Hermitian systems, where the
relevant dynamics is governed by a non-Hermitian operator whose spectral
degeneracies determine the stability and response of the system. The
identification of EPs in such settings, however, faces a fundamental
obstacle. The defining signature of an EP is eigenvector coalescence, yet in
classical platforms the accessible quantities are the physical
configurations themselves, the spin orientations, phases, amplitudes, or
currents that characterise the stationary state. The eigenvectors are not
physical observables: they describe infinitesimal perturbation directions
and are reconstructed only through a complete linearisation of the dynamics
around the equilibrium. In classical platforms, where the accessible
quantities are the physical configurations, the eigenvectors are therefore
neither directly measurable nor, in general, unambiguously extractable.

This obstacle defines the conceptual gap addressed in the present work. We
introduce classical observable coalescence as a diagnostic that
identifies EPs directly from the physical configurations of a nonlinear
classical system, without requiring eigenvector reconstruction. For two
equilibrium configurations $\mathbf X_1^\ast$ and $\mathbf X_2^\ast$,
characterised by physically meaningful variables such as spin orientations,
phases, or amplitudes, the observable separation is
$D_X=|\mathbf X_1^\ast-\mathbf X_2^\ast|$. Classical observable coalescence
corresponds to $D_X\rightarrow0$ as a control parameter approaches a critical
value. The central question is whether this observable-level phenomenon is
accompanied by the spectral and topological signatures that define an EP, and
whether it can distinguish an EP from a DP. The significance of this question
lies in its practical consequence: if classical observable coalescence tracks
EPs faithfully and produces no coalescence at DPs, then EPs become detectable
in classical many-body systems through the same quantities that are already
routinely measured.

To answer this question, we deliberately choose a model that has been studied
extensively in the quantum context: the one-dimensional non-Hermitian
transverse-field Ising chain with a staggered imaginary field, first analysed
by Castro-Alvarado and Fring \cite{castro2009spin} and subsequently
investigated in connection with $\mathcal{PT}$-symmetry breaking and quantum
phase transitions \cite{starkov2023quantum,lenke2021high,muldoon2023pt,
meisinger2012pt,wiersig2014enhancing}. This choice is methodological rather
than incidental. Because the quantum spectrum of this model is known, the
classical predictions developed here can be verified independently against
established results, ensuring that the proposed diagnostic is not an artefact
of the model but a reproducible feature of the non-Hermitian critical
structure. The Hamiltonian,
\begin{equation}
	H
	=
	J\sum_{\nu=1}^{N}\sigma_\nu^x\sigma_{\nu+1}^x
	+
	\sum_{\nu=1}^{N}\left[\eta+i(-1)^\nu\gamma\right]\sigma_\nu^z,
	\label{eq:H_intro}
\end{equation}
contains a nearest-neighbour Ising-like coupling $J$, a uniform real field
$\eta$, and a staggered imaginary field of strength $\gamma$ that alternates
between the two sublattices, balancing gain and loss across the chain and
endowing the Hamiltonian with $\mathcal{PT}$ symmetry. A concrete physical
realization of Eq.~\eqref{eq:H_intro}, in which the staggered gain and loss
are implemented through active electronic elements, is presented in
Appendix~\ref{app:platforms}.

The semiclassical description is constructed through a spin-coherent-state
representation, which places the analysis precisely at the boundary between
the quantum and classical regimes and provides the natural setting in which
classical observables emerge from the underlying quantum dynamics. Applying
the Heisenberg equations of motion to the Schwinger-boson representation of
the spin operators, and reducing the resulting operator equations through a
mean-field factorisation, we obtain a closed set of nonlinear flow equations
for the Bloch angles $\theta_\nu$ and relative phases $\phi_\nu$, together
with the global phase $\lambda_\nu$ that absorbs the non-Hermiticity and
preserves the norm of the state vector. One might object that the correspondence
between the classical and quantum descriptions is automatic, since both
derive from the same Hamiltonian but it is not the case. The classical
observables are obtained from a mean-field reduction of the Heisenberg
dynamics and are governed by a nonlinear system of flow equations; the
quantum spectrum is obtained from an exact Jordan-Wigner transformation and
Bogoliubov de-Gennes diagonalisation. These are different mathematical
objects derived through different procedures, and the fact that they encode
the same criticality is a non-trivial consistency result. It is precisely
this correspondence that allows classical observables to serve as proxies for
quantum EPs in platforms where the quantum spectrum is inaccessible.

The aims of this work are fourfold. First, we derive the equilibrium
configurations of the semiclassical flow equations and show that they
organise into two symmetry-related families whose branches lose their
distinguishability at family-dependent critical values of the non-Hermitian
parameter. Second, we evaluate the equilibrium energies of these branches and
show that they become degenerate at a family-independent critical scale,
exhibiting the characteristic square-root branch-point structure. Third, we
establish the topology of the degeneracy through analytic continuation of the
equilibrium branches into the complex $\gamma$ plane, examining the
encirclement of the branch point and the associated Riemann-sheet structure.
Fourth, we compare the classical results with the exact quantum spectrum and
demonstrate that the energy-level degeneracy occurs at the same critical
scale where the quantum spectral branches coalesce and their corresponding eigenvectors
become linearly dependent.

The remainder of this paper is organised as follows.
Section~\ref{sec:model} introduces the non-Hermitian spin model and develops
its semiclassical analysis, including the Schwinger-boson representation, the
mean-field flow equations, the equilibrium configurations, the observable
separation $D_X$, and the branch-resolved energies together with their
separation $D_E$. Section~\ref{sec:quantum} presents the exact quantum
treatment, including the Jordan-Wigner transformation, the Bogoliubov--de
Gennes diagonalisation, the dispersion relation, and the conditions for
spectral and eigenvector degeneracy. Section~\ref{sec:discussion} presents
the results and discussion, covering the classical observable coalescence,
the energy landscape, the comparison with the finite-size quantum spectrum,
the encirclement dynamics, the Riemann-sheet structure, and the distinction
between the exceptional point at $|\gamma|=|J|$ and the diabolic point at
$\gamma=0$. Section~\ref{sec:conclusions} summarises the principal findings
and discusses their implications.

\section{Non-Hermitian Spin Model and Semiclassical Analysis}
\label{sec:model}

We consider the one-dimensional non-Hermitian transverse-field Ising chain
with a staggered imaginary field,
\begin{equation}
	H
	=
	J\sum_{\nu=1}^{N}\sigma_\nu^x\sigma_{\nu+1}^x
	+
	\sum_{\nu=1}^{N}\left[\eta+i(-1)^\nu\gamma\right]\sigma_\nu^z,
	\label{eq:H}
\end{equation}
where $N$ is even, $J$ is the nearest-neighbour Ising-like coupling, $\eta$ is
the uniform real transverse field, and $\gamma$ controls the staggered
imaginary component. Periodic boundary conditions
$\sigma_{N+1}^{\nu}=\sigma_1^{\nu}$ are imposed. Odd sites belong to
sublattice $A$ and even sites to sublattice $B$, so that the local fields are
$g_A=\eta-i\gamma$ and $g_B=\eta+i\gamma$. The alternating sign of the
imaginary term produces opposite gain and loss on the two sublattices and
endows the Hamiltonian with $\mathcal{PT}$ symmetry, with parity interchanging
the sublattices and time reversal transforming $i\rightarrow-i$
\cite{castro2009spin,starkov2023quantum}. This model has been studied
extensively in the quantum context \cite{castro2009spin,starkov2023quantum,
lenke2021high,muldoon2023pt,meisinger2012pt}; we adopt it here as a
laboratory whose known quantum spectrum enables independent verification of
the classical predictions developed below. A concrete physical realization of
Eq.~\eqref{eq:H}, in which the staggered gain and loss are implemented
through active electronic elements, is presented in Appendix~\ref{app:platforms}.

To extract the classical observables, we employ the Schwinger-boson
representation of the spin operators,
\begin{equation}
	\sigma_\nu^z
	=
	2a_\nu^\dagger a_\nu-1,
	\qquad
	\sigma_\nu^x
	=
	a_\nu^\dagger b_\nu+b_\nu^\dagger a_\nu,
	\label{eq:SB}
\end{equation}
subject to the local constraint
$a_\nu^\dagger a_\nu+b_\nu^\dagger b_\nu=1$, and we follow the Heisenberg
dynamics of the bosonic operators,
$i\,d\hat{O}/dt=[\hat{O},\hat{H}]$. The resulting operator equations of
motion contain a nonlinear local effective field generated by the
nearest-neighbour interaction. To obtain a closed set of classical equations,
we associate with each lattice site a normalised Bloch state,
\begin{equation}
	|\psi_\nu\rangle
	=
	\cos\!\left(\frac{\theta_\nu}{2}\right)e^{i\lambda_\nu}|\psi_a\rangle
	+
	\sin\!\left(\frac{\theta_\nu}{2}\right)e^{i(\phi_\nu+\lambda_\nu)}
	|\psi_b\rangle,
	\label{eq:bloch}
\end{equation}
where $\theta_\nu$ is the polar angle, $\phi_\nu$ is the relative phase, and
$\lambda_\nu$ is the global phase that absorbs the non-Hermiticity and
preserves the norm of the state vector. Averaging the operator equations over
the Bloch state and factorising the resulting correlations in the mean-field
sense yields the classical flow equations
\begin{subequations}
	\label{eq:flow}
	\begin{align}
		\dot{\theta}_\nu
		&=
		2J\left[\sin(\theta_{\nu-1})\cos(\phi_{\nu-1})
		+
		\sin(\theta_{\nu+1})\cos(\phi_{\nu+1})\right]\sin(\phi_\nu)
		+
		2\gamma(-1)^\nu\sin(\theta_\nu),
		\\
		\dot{\phi}_\nu
		&=
		2J\left[\sin(\theta_{\nu-1})\cos(\phi_{\nu-1})
		+
		\sin(\theta_{\nu+1})\cos(\phi_{\nu+1})\right]\cot(\theta_\nu)
		\cos(\phi_\nu)
		-
		2\eta.
	\end{align}
\end{subequations}
The full derivation, including the treatment of the global phase
$\lambda_\nu$ and the statement of the mean-field approximations, is
presented in Appendix~\ref{app:meanfield}. The classical observables
$(\theta_\nu,\phi_\nu)$ are the physical variables of this nonlinear
dynamical system and are the quantities whose behaviour at criticality we
wish to characterise.

Exploiting the bipartite structure, we introduce the sublattice variables
$(\theta_A,\phi_A)$ for odd sites and $(\theta_B,\phi_B)$ for even sites. The
fixed points are obtained from the stationarity conditions
$\dot{\theta}_A=\dot{\theta}_B=\dot{\phi}_A=\dot{\phi}_B=0$. Restricting to
non-trivial solutions with $\sin\theta_A\neq0$ and $\sin\theta_B\neq0$, the
fixed-point equations admit a uniform polar angle $\theta_A=\theta_B\equiv
\theta$ together with the phase constraint
\begin{equation}
	\phi_A+\phi_B=n\pi,
	\qquad
	n\in\mathbb{Z}.
	\label{eq:phase_constraint}
\end{equation}
Two families of equilibrium configurations emerge, distinguished by the
parity of $n$. For the family with $n=2k$, one has
$\phi_B=-\phi_A \pmod{2\pi}$, and the equilibrium configurations are
\begin{equation}
	\phi_A
	=
	\frac{1}{2}\arcsin\!\left(-\frac{\gamma}{J}\right)+k\pi
	\quad\text{or}\quad
	\phi_A
	=
	\frac{\pi}{2}-\frac{1}{2}\arcsin\!\left(-\frac{\gamma}{J}\right)+k\pi,
	\label{eq:eq_F1}
\end{equation}
with
\begin{equation}
	\cos\theta
	=
	\frac{\eta}{2J\cos^2\phi_A},
	\qquad
	\phi_B=-\phi_A.
	\label{eq:theta_F1}
\end{equation}
For the family with $n=2k+1$, one has
$\phi_B=\pi-\phi_A \pmod{2\pi}$, and the equilibrium configurations are
\begin{equation}
	\phi_A
	=
	\frac{1}{2}\arcsin\!\left(\frac{\gamma}{J}\right)+k\pi
	\quad\text{or}\quad
	\phi_A
	=
	\frac{\pi}{2}-\frac{1}{2}\arcsin\!\left(\frac{\gamma}{J}\right)+k\pi,
	\label{eq:eq_F2}
\end{equation}
with
\begin{equation}
	\cos\theta
	=
	-\frac{\eta}{2J\cos^2\phi_A},
	\qquad
	\phi_B=\pi-\phi_A.
	\label{eq:theta_F2}
\end{equation}
Each family therefore contains two branches, labelled by the choice of the
phase solution in Eqs.~\eqref{eq:eq_F1} and~\eqref{eq:eq_F2}. We denote the
two branches of Family~1 by $F_1^{(\pm)}$ and those of Family~2 by
$F_2^{(\pm)}$.

To quantify the distinguishability of two equilibrium configurations, we
introduce the observable separation
\begin{equation}
	D_X
	=
	\left\|\mathbf X_1^\ast-\mathbf X_2^\ast\right\|,
	\qquad
	\mathbf X^\ast
	=
	(\theta_A,\theta_B,\phi_A,\phi_B),
	\label{eq:DX}
\end{equation}
where $\|\cdot\|$ denotes the Euclidean norm in the four-dimensional
observable space. The condition $D_X\rightarrow0$ signals that two
configurations that are physically distinct for generic parameters become
indistinguishable at a critical value of the control parameter. From
Eqs.~\eqref{eq:eq_F1}-\eqref{eq:theta_F2}, the two branches within each
family merge when the two phase solutions in Eq.~\eqref{eq:eq_F1}, and
correspondingly in Eq.~\eqref{eq:eq_F2}, coincide. This requires
\begin{equation}
	|\gamma|=|J|.
	\label{eq:merging}
\end{equation}
At this value the observable separation vanishes, $D_X=0$, and the two
distinct configurations of each family collapse onto a single fixed point.
The merging condition is independent of $\eta$, so that the non-Hermitian
parameter $\gamma$ alone controls the fusion of the classical configurations,
while $\eta$ determines the location of the merged configuration in the
$(\theta,\phi)$ plane. For $|\gamma|>|J|$, the fixed-point solutions cease to
be real, and no non-trivial equilibrium of the assumed form exists. The two
families merge at opposite critical values: $F_1^{(\pm)}$ merge at
$\gamma=-J$, and $F_2^{(\pm)}$ merge at $\gamma=+J$. This symmetry is a direct
consequence of the staggered gain-loss structure of the Hamiltonian and of
its $\mathcal{PT}$ symmetry, under which the transformation
$\gamma\rightarrow-\gamma$ interchanges the roles of the two sublattices.

The condition~\eqref{eq:merging} establishes that the classical
configurations lose their distinguishability at a well-defined value of the
non-Hermitian parameter. Whether this merging is accompanied by the spectral
and modal signatures of a non-Hermitian degeneracy cannot, however, be
determined from the equilibrium structure alone. It requires the analysis of
the equilibrium energies and, independently, of the exact quantum spectrum.
To complement the observable analysis, we therefore evaluate the expectation
value of the Hamiltonian~\eqref{eq:H} in the Bloch-state
representation~\eqref{eq:bloch}. The energy per site of the first family is
\begin{equation}
	E_{F_1}
	=
	\frac{J\pm\sqrt{J^2-\gamma^2}}{4}
	+
	\frac{\eta^2}{4\left(J\pm\sqrt{J^2-\gamma^2}\right)},
	\label{eq:E_F1}
\end{equation}
and that of the second family is
\begin{equation}
	E_{F_2}
	=
	-\frac{J\pm\sqrt{J^2-\gamma^2}}{4}
	-
	\frac{\eta^2}{4\left(J\pm\sqrt{J^2-\gamma^2}\right)},
	\label{eq:E_F2}
\end{equation}
so that $E_{F_2}=-E_{F_1}$. Both expressions depend on $\gamma$ through the
quantity
\begin{equation}
	S=\sqrt{J^2-\gamma^2},
	\label{eq:S}
\end{equation}
which controls the separation between the two branches within each family.
We define the corresponding energy separation
\begin{equation}
	D_E
	=
	\left|E_{F_i}^{(+)}-E_{F_i}^{(-)}\right|,
	\qquad
	i=1,2,
	\label{eq:DE}
\end{equation}
where $E_{F_i}^{(\pm)}$ denote the two branch energies of family $i$.

At $|\gamma|=|J|$, this separation vanishes, and the two energy branches of
each family become degenerate. It is important to distinguish this
energy-level degeneracy from the state-level merging established above. The
state configurations of Family~1 merge at $\gamma=-J$ and those of Family~2
at $\gamma=+J$, whereas the equilibrium energies of both families become
degenerate at $|\gamma|=|J|$. The energy-level degeneracy is therefore
family-independent and occurs at a single critical scale, in contrast to the
family-dependent merging of the configurations. This distinction is essential
when the semiclassical results are compared with the exact quantum
spectrum, since the latter is a statement about eigenenergies and the natural
point of comparison is accordingly the energy level, that is,
$|\gamma|=|J|$.

For $|\gamma|>|J|$, the energies acquire an imaginary
part, and the corresponding equilibrium configurations are no longer real.
The pair $(D_X,D_E)$ therefore provides two independent, physically
meaningful measures of the critical behaviour: $D_X$ probes the observable
level, and $D_E$ the energetic level. A genuine non-Hermitian degeneracy is
expected to manifest itself as the simultaneous vanishing of both quantities
at the same critical value of $\gamma$, and this is verified explicitly in
Sec.~\ref{sec:discussion}.

The appearance of the square-root quantity $S$ in
Eqs.~\eqref{eq:E_F1} and~\eqref{eq:E_F2} indicates that the equilibrium
branches are not single-valued functions of $\gamma$ in the neighbourhood of
$\gamma=\pm J$. Analytic continuation into the complex $\gamma$ plane
reveals a two-sheeted Riemann structure whose sheets meet at the branch
points $\gamma=\pm J$ and are exchanged upon encircling them. This
branch-point structure is a property of the semiclassical equilibrium
conditions alone and does not require any assumption about the quantum
spectrum. Its detailed topology, together with the encirclement dynamics of
the equilibrium branches, is examined in Sec.~\ref{sec:discussion}.

In summary, the semiclassical analysis of this section yields four results.
First, the equilibrium configurations of the non-Hermitian spin chain
organise into two symmetry-related families, each containing two branches,
with the two families merging at opposite values of the non-Hermitian
parameter, $\gamma=-J$ for Family~1 and $\gamma=+J$ for Family~2. Second, the
observable separation $D_X$ vanishes at $|\gamma|=|J|$, independently of
$\eta$, indicating the loss of distinguishability of the classical
configurations. Third, the branch-resolved equilibrium energies exhibit a
square-root dependence on $\gamma$ through $S=\sqrt{J^2-\gamma^2}$, and the
energy separation $D_E$ vanishes at the same critical value. Fourth, the
equilibrium branches possess a two-sheeted Riemann structure with branch
points at $\gamma=\pm J$, whose topology is analysed in
Sec.~\ref{sec:discussion}. The question of whether the merging observed at
$|\gamma|=|J|$ corresponds to a genuine exceptional point, in the sense of a
defective quantum operator with coalescing eigenvectors and non-trivial
branch-point topology, cannot be answered from the semiclassical analysis
alone. It requires the exact quantum treatment of the Hamiltonian~\eqref{eq:H},
which provides the spectrum and the eigenvectors whose coalescence defines a
non-Hermitian degeneracy. The quantum analysis is presented in
Sec.~\ref{sec:quantum}, and the confrontation of the classical and quantum
descriptions is carried out in Sec.~\ref{sec:discussion}.

\section{Quantum Analysis}
\label{sec:quantum}

The semiclassical analysis of Sec.~\ref{sec:model} established that the
equilibrium configurations of the non-Hermitian spin chain organise into two
symmetry-related families, each containing two branches, and that the two
branches within each family lose their distinguishability at a critical value
of the non-Hermitian parameter: the state configurations of Family~1 merge at
$\gamma=-J$, and those of Family~2 merge at $\gamma=+J$. At the level of the
equilibrium energies, however, the two branches of each family become
degenerate at the family-independent critical scale $|\gamma|=|J|$, and the
associated energy splitting depends on $\gamma$ through the square-root
quantity $S=\sqrt{J^2-\gamma^2}$, which is symmetric under
$\gamma\to-\gamma$. The energy-level degeneracy therefore occurs at a single
critical scale, in contrast to the family-dependent merging of the state
configurations. Since the quantum analysis that follows is a statement about
eigenenergies, the natural point of comparison with the semiclassical results
is the energy-level critical scale $|\gamma|=|J|$, and it is at this scale
that the quantum spectral branches are expected to become degenerate. The
exact quantum results derived here complement the mean-field derivation
presented in Appendix~\ref{app:meanfield} and provide the spectral benchmark
against which the circuit realisation of Appendix~\ref{app:platforms} can be
tested.

To determine whether these features correspond to a genuine coalescence of classical observables, and to establish the connection with the spectral properties of the parent quantum Hamiltonian, we now turn to an exact quantum treatment of the model. The Hamiltonian is that of Eq.~\eqref{eq:H},
\begin{equation*}
	H
	=
	J\sum_{\nu=1}^{N}\sigma_\nu^x\sigma_{\nu+1}^x
	+
	\sum_{\nu=1}^{N}\left[\eta+i(-1)^\nu\gamma\right]\sigma_\nu^z,
	\label{eq:H_quantum}
\end{equation*}
and our aim is to diagonalise it, determine its eigenenergies and eigenvectors, and identify the conditions under which its spectral branches become degenerate.

A direct diagonalisation of the full many-body Hamiltonian becomes rapidly demanding with increasing system size because the Hilbert-space dimension grows exponentially as $2^N$. A more efficient description is possible when the spin model can be mapped onto a quadratic fermionic Hamiltonian. For the transverse-field Ising model, this reduction is achieved exactly through the Jordan-Wigner transformation, which maps the spin operators onto fermionic creation and annihilation operators, converting the nearest-neighbour spin interaction into fermionic hopping and pairing terms. The staggered complex field is simultaneously transformed into a complex site-dependent fermionic potential. The resulting fermionic Hamiltonian retains the full information of the original spin model while allowing the many-body problem to be expressed in terms of bilinear fermionic operators.

Because the non-Hermitian field alternates from site to site, the transformed system possesses a natural two-sublattice structure. We therefore introduce separate fermionic operators for the two sublattices and subsequently employ a Fourier transformation to express the Hamiltonian in momentum space. This procedure separates the lattice problem into independent momentum sectors. Because the Jordan-Wigner transformation generates anomalous terms of the form $c_\nu^\dagger c_{\nu+1}^\dagger$ and $c_\nu c_{\nu+1}$ in addition to the usual hopping terms, the momentum-space Hamiltonian is naturally written in the Bogoliubov-de Gennes (BdG) formalism \cite{Jipdi2026TopologicallyProtected}. A Nambu spinor containing particle and hole degrees of freedom is introduced, leading to a finite matrix for each momentum $k$. The quasiparticle energies are then obtained from the eigenvalues of this BdG matrix.

Introducing fermionic operators $c_\nu$ and $c_\nu^\dagger$ that satisfy the canonical anticommutation relations
\begin{equation}
	\{c_\nu,c_j^\dagger\}=\delta_{\nu j},
	\qquad
	\{c_\nu,c_j\}=0,
	\qquad
	\{c_\nu^\dagger,c_j^\dagger\}=0,
	\label{eq:fermion_anticomm}
\end{equation}
The spin raising and lowering operators are written as
\begin{subequations}
	\label{eq:JW}
	\begin{align}
		\sigma_\nu^+
		&=
		\sigma_\nu^x+i\sigma_\nu^y
		=
		\left(\prod_{l=1}^{\nu-1}\sigma_l^z\right)c_\nu^\dagger,
		\\
		\sigma_\nu^-
		&=
		\sigma_\nu^x-i\sigma_\nu^y
		=
		\left(\prod_{l=1}^{\nu-1}\sigma_l^z\right)c_\nu,
		\\
		\sigma_\nu^z
		&=
		2c_\nu^\dagger c_\nu-1.
	\end{align}
\end{subequations}
Substitution of Eq.~\eqref{eq:JW} into the Hamiltonian~\eqref{eq:H_quantum} yields the quadratic fermionic Hamiltonian
\begin{equation}
	\begin{aligned}
		H
		={}&
		J\sum_{\nu=1}^{N-1}
		\left(
		c_\nu^\dagger c_{\nu+1}
		+
		c_{\nu+1}^\dagger c_\nu
		+
		c_\nu^\dagger c_{\nu+1}^\dagger
		-
		c_\nu c_{\nu+1}
		\right)
		\\
		&+
		2\sum_{\nu=1}^{N}
		\left[\eta+i\gamma(-1)^\nu\right]
		c_\nu^\dagger c_\nu
		+
		\sum_{\nu=1}^{N}
		\left[\eta+i\gamma(-1)^\nu\right],
	\end{aligned}
	\label{eq:H_fermion}
\end{equation}
which contains hopping, pairing, and complex on-site terms. The quadratic structure of Eq.~\eqref{eq:H_fermion} is the key feature that makes the subsequent Bogoliubov-de Gennes analysis possible.

We take an even number of sites, $N=2M$, so that gain and loss are balanced across the lattice. The alternating non-Hermitian field makes it convenient to introduce two sublattices through
\begin{equation}
	A_m=c_{2m},
	\qquad
	B_m=c_{2m-1},
	\qquad
	m=1,\ldots,M.
	\label{eq:sublattice_fermion}
\end{equation}
In terms of these operators, the fermionic Hamiltonian becomes
\begin{equation}
	\begin{aligned}
		H_{\mathrm{F}}
		={}&
		J\sum_{m=1}^{M}
		\left[
		B_{2m-1}^\dagger A_{2m}
		+
		A_{2m}^\dagger B_{2m-1}
		+
		A_{2m}^\dagger B_{2m+1}
		+
		B_{2m+1}^\dagger A_{2m}
		\right]
		\\
		&+
		J\sum_{m=1}^{M}
		\left[
		B_{2m-1}^\dagger A_{2m}^\dagger
		-
		B_{2m-1}A_{2m}
		+
		A_{2m}^\dagger B_{2m+1}^\dagger
		-
		A_{2m}B_{2m+1}
		\right]
		\\
		&+
		2\sum_{m=1}^{M}
		\left[
		(\eta+i\gamma)A_{2m}^\dagger A_{2m}
		+
		(\eta-i\gamma)B_{2m-1}^\dagger B_{2m-1}
		\right]
		+
		N\eta.
	\end{aligned}
	\label{eq:H_sublattice}
\end{equation}
We then introduce the Fourier transforms
\begin{equation}
	A_{2m}
	=
	\frac{1}{\sqrt{M}}\sum_k e^{i2km}A_k,
	\qquad
	B_{2m-1}
	=
	\frac{1}{\sqrt{M}}\sum_k e^{ik(2m-1)}B_k,
	\label{eq:Fourier}
\end{equation}
where the momentum sum runs over the allowed values $k=2\pi n/M$ with $n=0,\ldots,M-1$. Substitution of Eq.~\eqref{eq:Fourier} into Eq.~\eqref{eq:H_sublattice} gives the momentum-space Hamiltonian
\begin{equation}
	\begin{aligned}
		H_{\mathrm{F}}
		={}&
		2J\sum_k \cos k
		\left(
		A_k^\dagger B_k
		+
		B_k^\dagger A_k
		\right)
		\\
		&+
		2iJ\sum_k \sin k
		\left(
		A_k^\dagger B_{-k}^\dagger
		+
		A_k B_{-k}
		\right)
		\\
		&+
		2\sum_k
		\left[
		(\eta+i\gamma)A_k^\dagger A_k
		+
		(\eta-i\gamma)B_k^\dagger B_k
		\right]
		+
		N\eta.
	\end{aligned}
	\label{eq:H_momentum}
\end{equation}
The anomalous pairing terms couple particles at momentum $k$ to holes at momentum $-k$. We therefore introduce the four-component Nambu spinor
\begin{equation}
	\Psi_k
	=
	\begin{pmatrix}
		A_k \\
		B_k \\
		A_{-k}^\dagger \\
		B_{-k}^\dagger
	\end{pmatrix},
	\label{eq:Nambu}
\end{equation}
in terms of which the Hamiltonian assumes the Bogoliubov-de Gennes form
\begin{equation}
	H_{\mathrm{F}}
	=
	\frac{1}{2}\sum_k \Psi_k^\dagger \mathcal{H}_{\mathrm{BdG}}(k)\Psi_k
	+
	E_{\mathrm{c}},
	\label{eq:BdG_form}
\end{equation}
where $E_{\mathrm{c}}$ is a constant. The corresponding BdG matrix is
\begin{equation}
	\mathcal{H}_{\mathrm{BdG}}(k)
	=
	\begin{pmatrix}
		2(\eta+i\gamma) & 2J\cos k & 0 & 2iJ\sin k \\[4pt]
		2J\cos k & 2(\eta-i\gamma) & 2iJ\sin k & 0 \\[4pt]
		0 & -2iJ\sin k & -2(\eta-i\gamma) & -2J\cos k \\[4pt]
		-2iJ\sin k & 0 & -2J\cos k & -2(\eta+i\gamma)
	\end{pmatrix}.
	\label{eq:BdG_matrix}
\end{equation}
The diagonal blocks contain the particle and hole sectors, while the off-diagonal blocks describe the anomalous particle-hole coupling generated by the $XX$ interaction. The quasiparticle energies are obtained from the eigenvalues of $\mathcal{H}_{\mathrm{BdG}}(k)$. Diagonalisation yields the dispersion relation
\begin{equation}
	E_{k,\pm}
	=
	\pm 2
	\sqrt{
		J^2+\eta^2-\gamma^2
		\pm
		2\sqrt{
			J^2(\eta^2+\gamma^2)\cos^2 k
			-
			\gamma^2(J^2+\eta^2)
		}
	},
	\label{eq:dispersion}
\end{equation}
so that the complete BdG spectrum consists of four eigenvalues for each momentum $k$, appearing in $\pm$ pairs.

To trace the conditions under which the spectral branches become degenerate, we analyse non-Hermitian degeneracies at which eigenvalues of the BdG matrix become identical simultaneously with the coalescence of their corresponding eigenvectors. In contrast to ordinary degeneracies in Hermitian systems, such a degeneracy renders the Hamiltonian non-diagonalizable and is therefore associated with a defective eigenvalue. The energy branches become degenerate when the discriminant associated with the inner square root in Eq.~\eqref{eq:dispersion} vanishes,
\begin{equation}
	J^2(\eta^2+\gamma^2)\cos^2 k
	-
	\gamma^2(J^2+\eta^2)
	=
	0.
	\label{eq:discriminant}
\end{equation}
Hence, the momentum at which the spectral branches become degenerate satisfies
\begin{equation}
	\cos^2 k_{\mathrm{EP}}
	=
	\frac{
		\gamma^2(J^2+\eta^2)
	}{
		J^2(\eta^2+\gamma^2)
	}.
	\label{eq:k_EP}
\end{equation}
This relation provides the degeneracy supersurface that gives the momentum-space location of the spectral degeneracies for a specified set of model parameters. For a finite periodic chain, the allowed momenta are quantised according to $k_n=\pi n/M$ with $n\in\{0,\ldots,M-1\}$, and each value of $k_{\mathrm{EP}}$ corresponds to a set of degeneracies in real space, which allows all degeneracies of a finite system to be identified.

The boundary of the degeneracy supersurface is obtained for $k=0,\pi$, for which $\cos^2 k=1$. Substituting into Eq.~\eqref{eq:k_EP} gives
\begin{equation}
	\gamma^2(J^2+\eta^2)
	=
	J^2(\eta^2+\gamma^2),
	\label{eq:boundary_condition}
\end{equation}
which reduces to
\begin{equation}
	|\gamma|=|J|.
	\label{eq:gamma_J}
\end{equation}
The value $|\gamma|=|J|$ therefore separates the parameter region in which the degeneracy condition can be realised at real crystal momentum from the region in which no real momentum satisfies Eq.~\eqref{eq:k_EP}. For the high-symmetry momenta $k=0$ and $k=\pi$, the BdG Hamiltonian simplifies considerably: the particle and hole sectors become decoupled with respect to the anomalous contribution, and the eigenvalues reduce to
\begin{equation}
	E_{p,\pm}
	=
	\pm 2\eta
	\pm
	2\sqrt{J^2-\gamma^2}.
	\label{eq:E_high_sym}
\end{equation}
The two particle-sector eigenvalues become degenerate when
$J^2-\gamma^2=0$
that is, when $|\gamma|=|J|$. At this point, the square-root splitting characteristic of the non-Hermitian two-level problem vanishes.

To make the structure of the degeneracy explicit, we introduce the angle
\begin{equation}
	\phi
	=
	\arctan\!\left(
	\frac{\gamma}{\sqrt{J^2-\gamma^2}}
	\right).
	\label{eq:phi}
\end{equation}
The particle-sector eigenvectors may then be written, up to an arbitrary normalisation and overall phase, as
\begin{equation}
	\psi_{p,\pm}
	=
	\frac{1}{\sqrt{2}}
	\begin{pmatrix}
		\pm e^{\mp i\phi} \\
		1 \\
		0 \\
		0
	\end{pmatrix},
	\label{eq:eigenvector_particle}
\end{equation}
while the corresponding hole-sector states are
\begin{equation}
	\psi_{h,\pm}
	=
	\frac{1}{\sqrt{2}}
	\begin{pmatrix}
		0 \\
		0 \\
		\pm e^{\mp i\phi} \\
		1
	\end{pmatrix}.
	\label{eq:eigenvector_hole}
\end{equation}
As $|\gamma|\to|J|$, the quantity $\sqrt{J^2-\gamma^2}$ vanishes and the two eigenvectors within each sector become identical. This eigenvector degeneracy, together with the corresponding eigenvalue degeneracy, constitutes the defining signature of a non-Hermitian spectral degeneracy at the high-symmetry momenta.

The quantum analysis of this section therefore provides three results that
parallel and complement those of Sec.~\ref{sec:model}. First, the exact BdG
spectrum exhibits a square-root branch-point singularity at $|\gamma|=|J|$,
in direct correspondence with the square-root dependence of the classical
equilibrium energies. Second, the quantum spectral separation vanishes
at $|\gamma|=|J|$, which is precisely the same critical scale at which the
classical equilibrium energy separation $D_E$ vanishes. The comparison at the
energy level is therefore exact: both descriptions exhibit the same
family-independent critical scale and the same square-root branch-point
structure. Third, the BdG eigenvectors become linearly dependent at
$|\gamma|=|J|$, establishing that the degeneracy is defective rather than
ordinary. The relation between these quantum spectral features and the semiclassical equilibrium results of Sec.~\ref{sec:model}, together with their physical interpretation, is discussed in Sec.~\ref{sec:discussion}.

\section{Results and Discussions}
\label{sec:discussion}

Having established the semiclassical equilibrium structure in
Sec.~\ref{sec:model} and the exact quantum spectral structure in
Sec.~\ref{sec:quantum}, we now confront the two descriptions. The purpose of
this section is fourfold. First, we follow the equilibrium branches directly
through the physical classical observables and identify the parameter values
at which distinct configurations lose their distinguishability. Second, we
examine the corresponding equilibrium energies and establish the critical
scale at which they become degenerate. Third, we use the exact quantum
spectrum to determine whether the observed degeneracy is an ordinary crossing
or a genuine exceptional point, and we verify its branch-point topology
through encirclement and Riemann-sheet analysis. Fourth, we distinguish the
exceptional point at $|\gamma|=|J|$ from the diabolic point at $\gamma=0$ and
assess the discriminating power of the observable-level diagnostic. We
therefore analyse the classical observables, examine the equilibrium
energies, confront the classical degeneracy with the exact quantum spectrum,
test the branch-point topology by encircling the degeneracy in the complex
$\gamma$ plane, and construct the Riemann-sheet structure that characterises
the exceptional point.

\subsection{Classical equilibrium branches and their fusion}
\label{subsec:classical_fusion}

\begin{figure}[htbp]
	\centering
	\includegraphics[height=11cm,width=17.0cm]{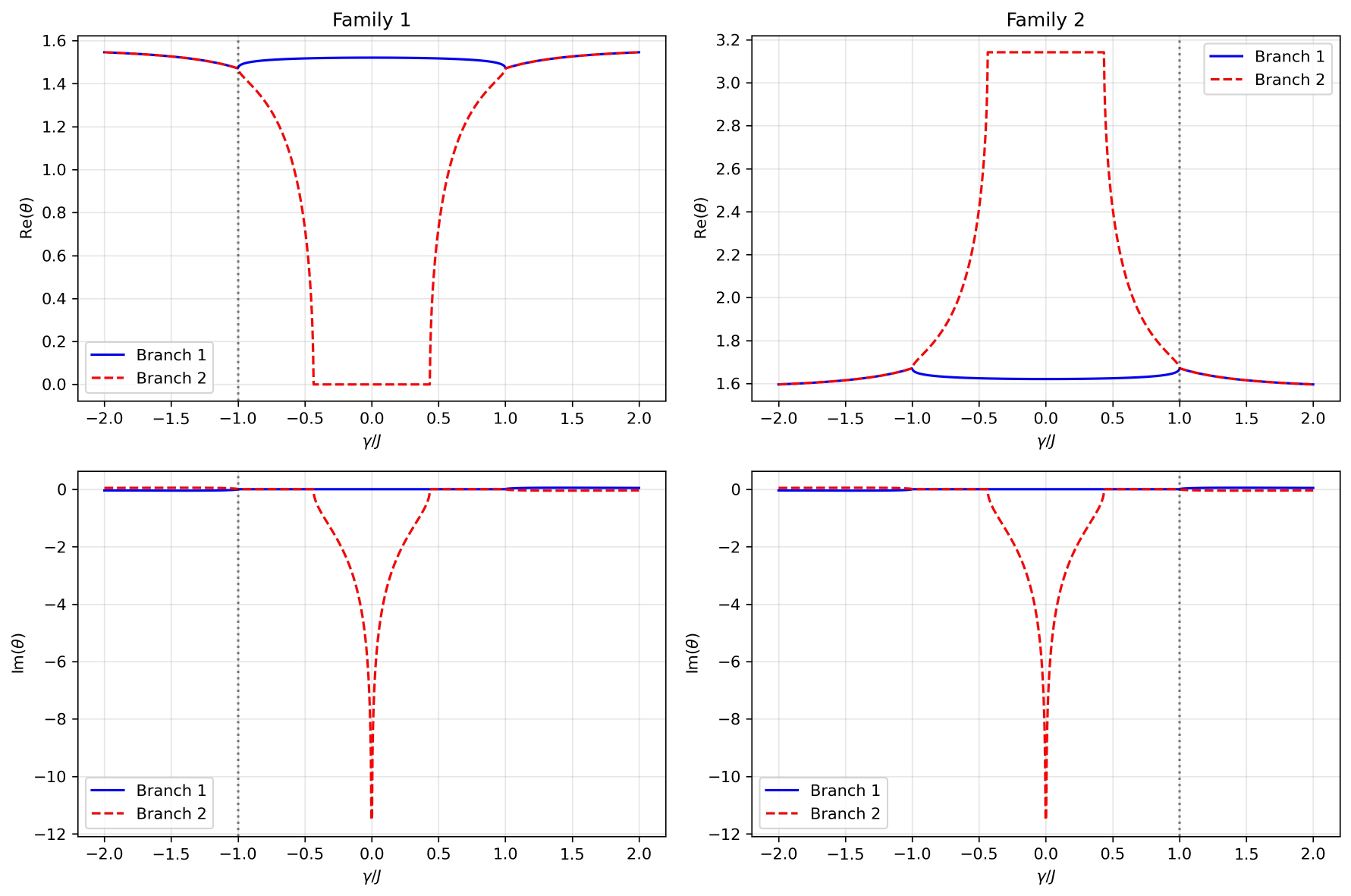}
	\caption{
		Fusion of the classical equilibrium configurations via the polar
		observable for Families~1 and~2 as a function of the non-Hermitian
		control parameter $\gamma/J$ at $\eta/J=0.1$. Panels (a) and (b)
		display the real part of the polar variable $\theta$ for Families~1
		and~2, respectively; panels (c) and (d) display the corresponding
		imaginary parts. The two equilibrium branches of each family merge at the family-dependent values $\gamma/J=-1$ for Family~1 and $\gamma/J=+1$ for Family~2.
	}
	\label{fig:classical_coalescence_theta}
\end{figure}

\begin{figure}[htbp]
	\centering
	\includegraphics[height=11cm,width=17.0cm]{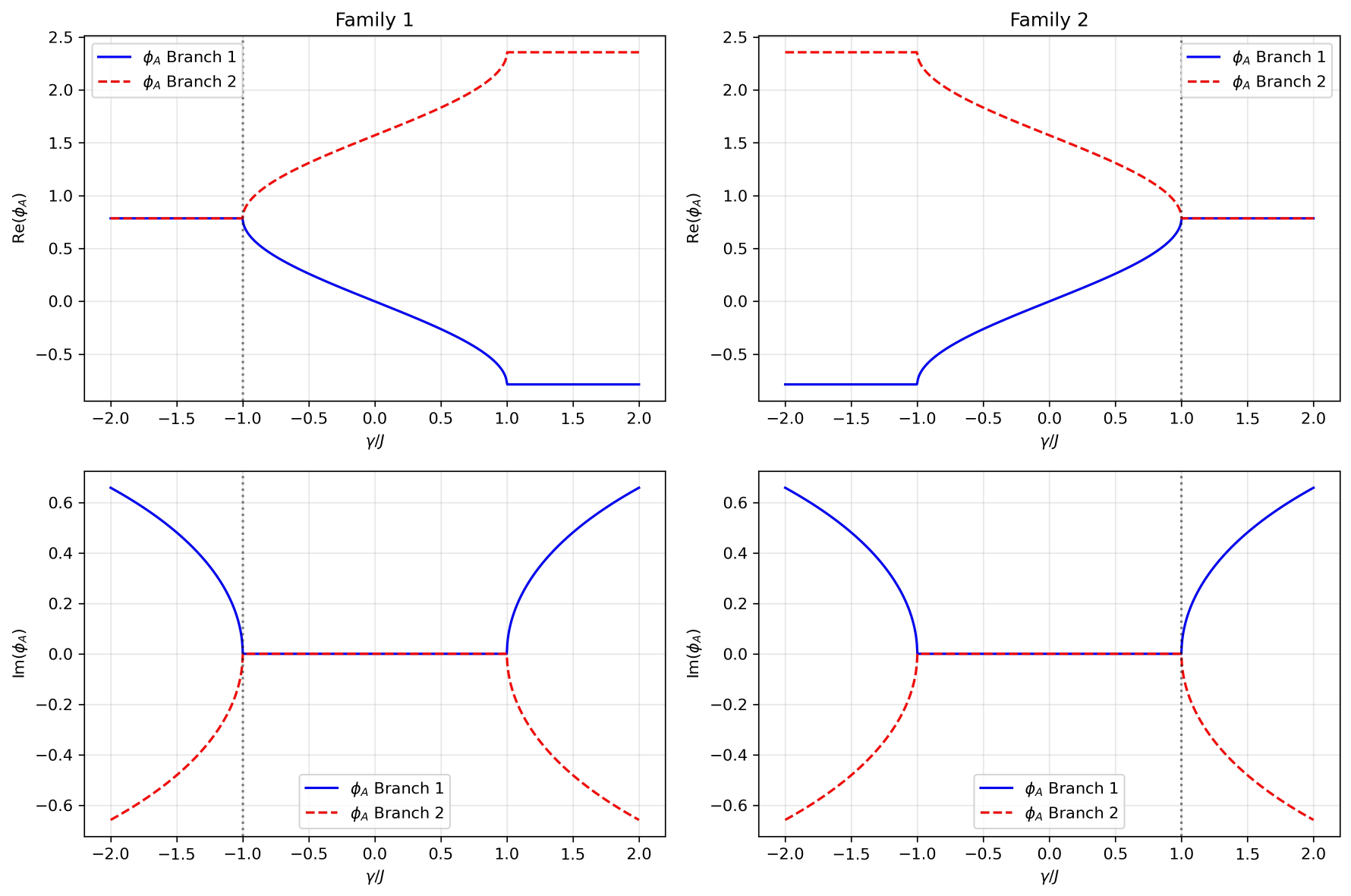}
	\caption{
		Fusion of the classical equilibrium configurations via the
		relative-phase observable for Families~1 and~2 as a function of
		$\gamma/J$ at $\eta/J=0.1$. Panels (a) and (b) display the real part
		of $\phi$ for Families~1 and~2, respectively; panels (c) and (d)
		display the corresponding imaginary parts. The two branches of each
		family merge at $\gamma/J=-1$ for Family~1 and $\gamma/J=+1$ for
		Family~2. Beyond these points, the analytically continued branches
		acquire nonzero imaginary components.
	}
	\label{fig:classical_coalescence_phi}
\end{figure}

As demonstrated in Sec.~\ref{sec:model}, the fixed-point analysis of the
mean-field equations yields two families of equilibrium configurations,
distinguished by the relative phase and characterised by
Eqs.~\eqref{eq:eq_F1} and~\eqref{eq:eq_F2}. We present the real and imaginary
parts of the equilibrium polar and azimuthal variables for the two families
in Figs.~\ref{fig:classical_coalescence_theta}
and~\ref{fig:classical_coalescence_phi}.

\begin{figure}[htbp]
	\centering
	\includegraphics[height=6.0cm,width=17.0cm]{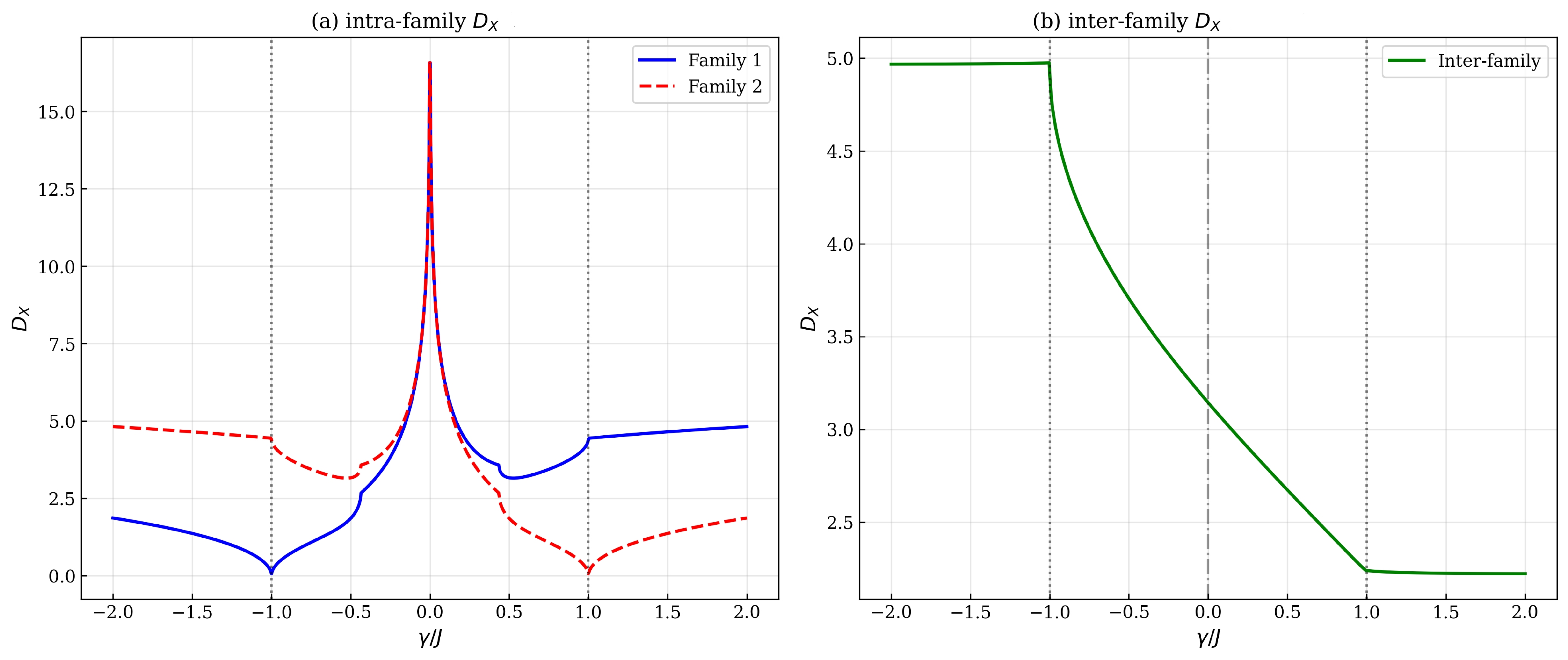}
	\caption{
		Observable separation $D_X$ between the two branches within each
		family as a function of $\gamma/J$ at $\eta/J=0.1$. Panel (a)
		displays the intra-family separation, with the solid curve
		corresponding to Family~1 and the dashed curve to Family~2; both
		curves vanish at the family-dependent critical values
		$\gamma/J=-1$ and $\gamma/J=+1$, signalling the merging of the
		classical state configurations. Panel (b) displays the inter-family
		separation between one branch of Family~1 and one branch of
		Family~2; the separation decreases but saturates and never vanishes,
		indicating that configurations belonging to different families
		remain distinguishable throughout the parameter range.
	}
	\label{fig:DX}
\end{figure}

Figure~\ref{fig:classical_coalescence_theta} shows the variation of the polar
angle with $\gamma/J$, while Fig.~\ref{fig:classical_coalescence_phi}
displays the corresponding variation of the relative phase. The behaviour of
the two observables is markedly different, and this difference is physically
informative.

The real part of the polar variable $\theta$, shown in panels (a) and (b) of
Fig.~\ref{fig:classical_coalescence_theta}, remains essentially superimposed
throughout the parameter range $|\gamma|>J$, so that the polar components of
the two branches carry no distinguishing information in this region. The
corresponding imaginary parts, shown in panels (c) and (d), vanish for
$|\gamma|>J$ and become nonzero out of this interval, merging at the
family-dependent critical values $\gamma=-J$ and $\gamma=+J$. For $|\gamma|<J$, the two branches separates and become distinguishable as seen in the four panels.   This indicates
that, for the parameter slice considered here, the loss of distinguishability
between the two equilibrium configurations is encoded  in  both real and 
complex extension of the polar variable rather than in its imaginary part only, a
characteristic feature of non-Hermitian systems in which physically
meaningful observables may acquire complex components once the system leaves
the regime of real spectra \cite{ashida2020non,heiss2012physics,
moiseyev2011non}.

The panels of Fig.~\ref{fig:classical_coalescence_phi} provide the decisive
and more precise signature of the fusion. For $|\gamma|<J$, the real parts of
the two phase branches are clearly separated, indicating two distinguishable
equilibrium configurations. As $\gamma/J$ increases toward the critical
value, the separation decreases continuously and vanishes at $\gamma=-J$ for
Family~1 and at $\gamma=+J$ for Family~2. At these points, the two branches
of each family satisfy $\phi_{F_i}=\phi_{F_i}'$, so that the two previously
distinct stationary configurations become indistinguishable in the space of
physical observables. This is precisely the phenomenon that we refer to as
classical observable coalescence. The imaginary parts of $\phi$, shown in
panels (c) and (d), reveal the complementary behaviour: for $|\gamma|<J$, the
equilibrium phases are real and $\operatorname{Im}(\phi)=0$; once
$|\gamma|>J$, the square-root quantity $S=\sqrt{J^2-\gamma^2}$ becomes
imaginary, and the analytically continued branches acquire nonzero imaginary
components. The fusion observed on the real control-parameter axis is
therefore the real-axis manifestation of a larger complex branch structure.

To quantify the loss of distinguishability directly, we display in
Fig.~\ref{fig:DX} the observable separation $D_X$ defined in
Eq.~\eqref{eq:DX}, evaluated between the two branches within each family in
panel (a) and between two branches of different families in panel (b). In
panel (a), the solid curve corresponds to Family~1 and the dashed curve to
Family~2. In both cases, $D_X$ decreases monotonically as the respective
critical value is approached, and vanishes exactly at the family-dependent
merging points $\gamma=-J$ (Family~1) and $\gamma=+J$ (Family~2). The
vanishing of $D_X$ is sharp and is not accompanied by any discontinuity in
the slope, consistent with the square-root dependence of the underlying
branch separation. In panel (b), by contrast, the inter-family separation
decreases but saturates and never vanishes: configurations belonging to
different families remain distinguishable throughout the parameter range.
This figure provides the primary observable-level evidence for the merging
of the classical configurations.

The two families exhibit the same physical mechanism of classical observable
coalescence but at opposite values of the non-Hermitian control parameter.
The symmetry between the two families is significant: it demonstrates that
they are not independent accidental solutions but are related by the
staggered gain-loss structure of the model. The transformation
$\gamma\rightarrow-\gamma$ interchanges the locations of the two critical
points while preserving the qualitative branch geometry, so that the
classical equilibrium manifold possesses a symmetry-related pair of
observable fusions. This symmetry is a direct consequence of the
$\mathcal{PT}$-symmetric structure of the Hamiltonian~\eqref{eq:H}, under
which parity exchanges the two sublattices and time reversal inverts the sign
of $\gamma$.

Two remarks on the interpretation of these results are in order. First, the
merging of the state configurations at $\gamma=-J$ (Family~1) and
$\gamma=+J$ (Family~2) is a statement about the physical observables
$(\theta,\phi)$ and is therefore family-dependent. Second, the appearance of
complex-valued observables beyond the real-solution interval is physically
significant: rather than terminating abruptly at the critical point, the
analytical branches can be continued into the complex domain, revealing that
the real classical configuration becomes part of a multivalued structure
once the non-Hermitian control parameter exceeds the real-solution boundary.
This behaviour is reminiscent of the complex energy surfaces that arise in
the description of non-Hermitian phase transitions and provides a natural
geometric language for the observed fusion \cite{heiss2012physics,
berry2004physics,miri2019exceptional}. It is now of interest to determine
whether other physical observables, in particular the equilibrium energy,
exhibit the same non-Hermitian phenomenon.

\subsection{Energy landscape of the equilibrium branches}
\label{subsec:energy_landscape}

\begin{figure}[htbp]
	\centering
\includegraphics[width=17cm,height=11cm]{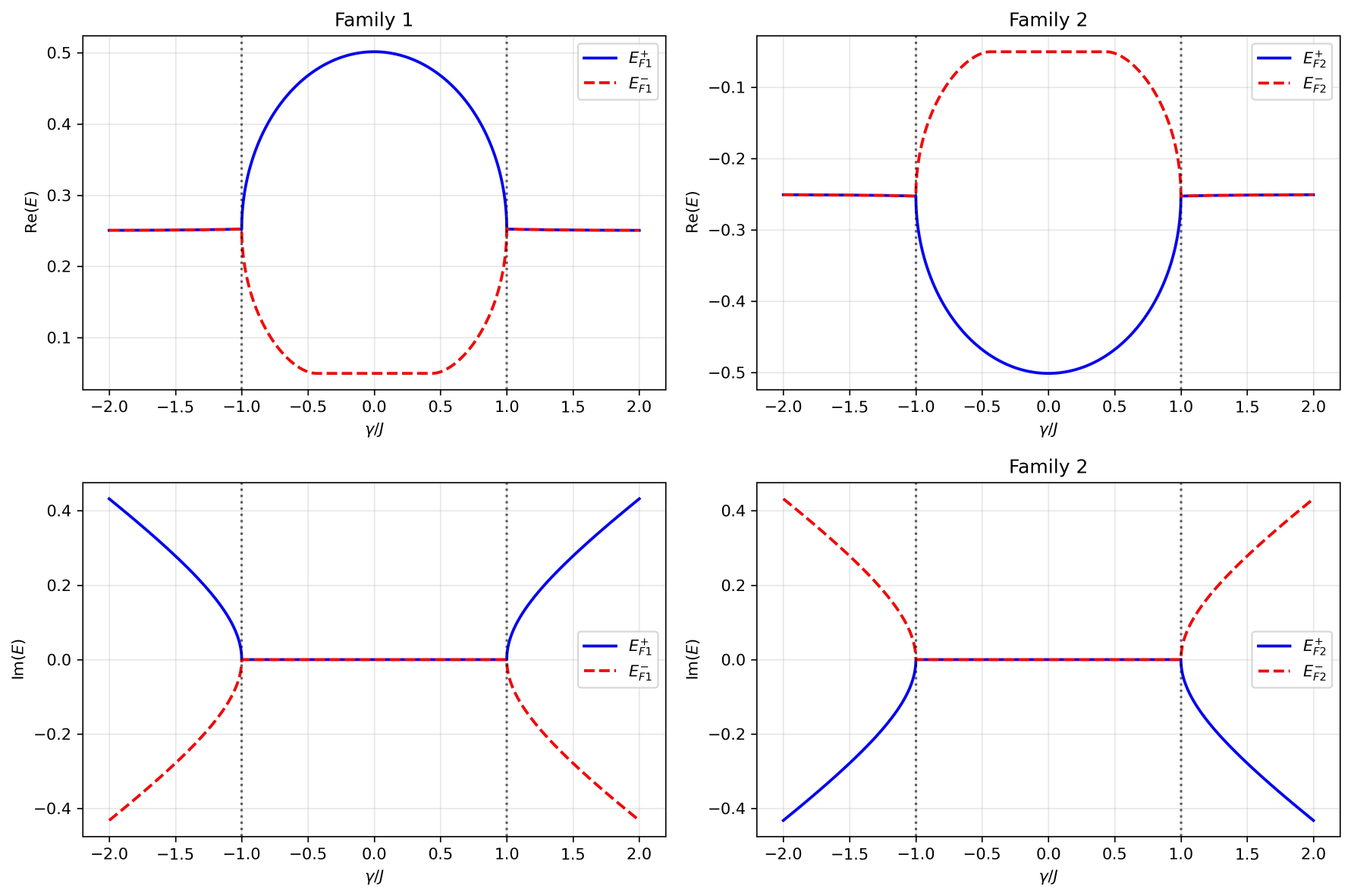}
	\caption{
		Energy landscape as a function of the non-Hermitian parameter
		$\gamma/J$ at $\eta/J=0.1$. (a) Energy branches of Family~1.
		(b) Energy branches of Family~2. The two real branches of each
		family approach one another and become degenerate at $|\gamma/J|=1$.
	}
	\label{fig:energy_land}
\end{figure}

\begin{figure}[htbp]
	\centering
	\includegraphics[height=6.0cm,width=17.0cm]{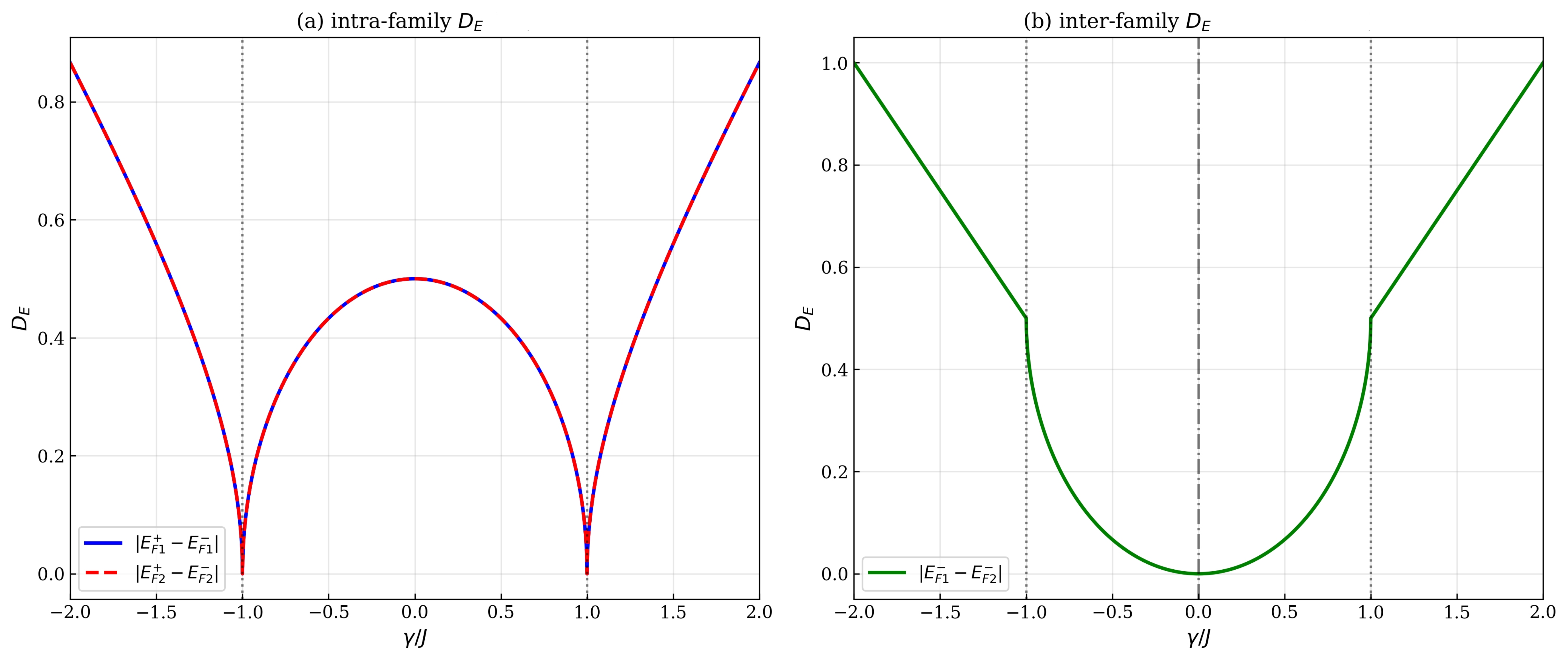}
	\caption{
		Energy separation $D_E$ as a function of $\gamma/J$ at
		$\eta/J=0.1$. Panel (a) shows the intra-family separation, with the
		solid curve corresponding to Family~1 and the dashed curve to
		Family~2; both curves vanish at the family-independent critical
		scale $|\gamma/J|=1$. Panel (b) shows the inter-family separation
		$|E_{F1}^{-}-E_{F2}^{-}|$, which vanishes at $\gamma=0$ and
		therefore identifies a potential degeneracy point at the origin of
		the parameter space.
	}
	\label{fig:DE}
\end{figure}

\begin{figure}[htbp]
	\centering
	\includegraphics[width=17cm,height=6cm]{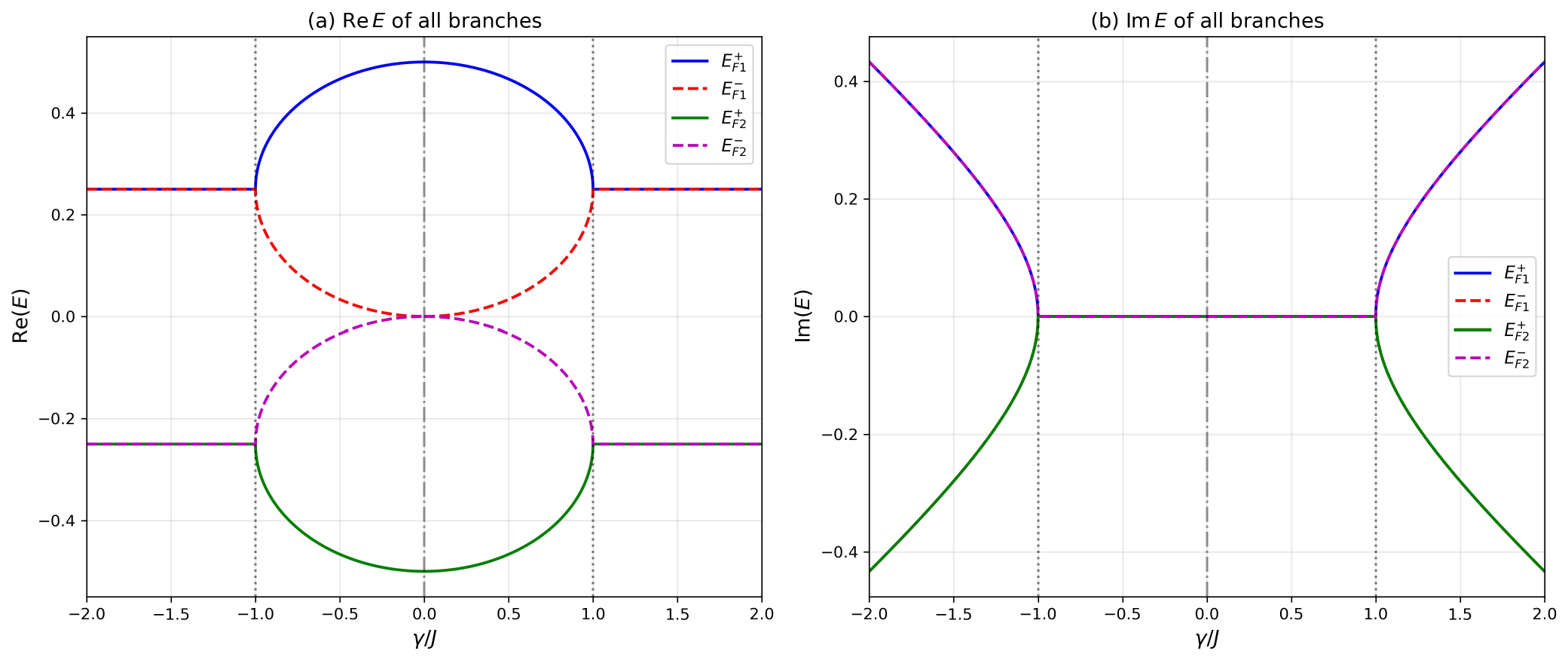}
	\caption{
		Energy branches of Families~1 and~2 displayed together for
		$\eta/J=0.0$. The branches of the two families cross at $\gamma=0$.
		Because the Hamiltonian at $\gamma=0$ is Hermitian, this crossing is
		an accidental degeneracy (a diabolic point) rather than a potential
		exceptional point.
	}
	\label{fig:riemann_combined}
\end{figure}

Having established the merging of the classical equilibrium configurations at
the family-dependent values $\gamma=\pm J$, we now examine its manifestation
in the energy landscape. Using the branch-resolved energies derived in
Sec.~\ref{sec:model}, Eqs.~\eqref{eq:E_F1} and~\eqref{eq:E_F2}, we plot the
energy branches of Families~1 and~2 in Fig.~\ref{fig:energy_land}, Energy separation $D_E$ as a function of $\gamma/J$ in Fig.~\ref{fig:DE} and the combined branches for $\eta/J=0$  in Fig.~\ref{fig:riemann_combined}   .

The separation between the two branches is controlled directly by the
quantity $\sqrt{J^2-\gamma^2}$, which appears in both energy expressions.
Figure~\ref{fig:energy_land} shows that the two real energy branches of each
family approach one another and meet at $|\gamma/J|=1$, becoming identical at
both $\gamma/J=+1$ and $\gamma/J=-1$. The energetic description therefore
reproduces the same critical scale that was identified from the quantum
analysis, and this agreement is the more significant because the classical
state configurations merge at family-dependent values while the energies
become degenerate at a family-independent one.

To make this distinction explicit, we display in Fig.~\ref{fig:DE} the
energy separation $D_E$ defined in Eq.~\eqref{eq:DE}, evaluated between the
two branches within each family in panel (a) and between the two branches
$E_{F1}^{-}$ and $E_{F2}^{-}$ of different families in panel (b). The solid
and dashed curves of panel (a) vanish at the same critical scale
$|\gamma/J|=1$, reflecting the fact that the energy separation depends on
$\gamma$ only through the symmetric quantity $\sqrt{J^2-\gamma^2}$, so that
the energy-level degeneracy is family-independent. Panel (b) reveals an
additional feature: the inter-family energy separation vanishes at
$\gamma=0$, indicating that the two branches belonging to different families
become degenerate at the origin of the parameter space. This is a potential
additional degeneracy point, and its character must be established before any
identification can be made.

Moreover, Fig.~\ref{fig:riemann_combined} shows that the branches of each
family meet at $\gamma=\pm J$, as discussed above. When the two families are
considered together, however, an additional crossing occurs at $\gamma=0$.
At the classical level, this point could in principle be interpreted as
another degeneracy point to be investigated, but such an interpretation would
violate basic physical requirements. At $\gamma=0$, the
Hamiltonian~\eqref{eq:H} is Hermitian, and the notion of exceptional dynamics
is not applicable to a Hermitian operator, whose eigenvectors can always be
chosen orthogonal. This elementary observation must, however, be supported by
the detailed physics of the problem, and this is precisely what the
observable-separation and encirclement analyses provide. As we have
established in Fig.~\ref{fig:DX}(b), the inter-family observable separation
$D_X$ does not vanish at $\gamma=0$, so the degeneracy in energy is not
accompanied by a fusion of the classical configurations. The combined
evidence identifies the degeneracy at $\gamma=0$ as a diabolic point rather
than a potential exceptional point.

It is important to emphasise that the degeneracy of the energy branches is,
at this stage, an observation at the level of the classical equilibrium
energies. A degeneracy of energies alone does not establish that the
underlying dynamics is defective, nor does it distinguish an ordinary
crossing from a genuine exceptional point. To settle this question, and to
determine whether the observed classical observable coalescence is a
manifestation of an exceptional point, we must turn to the exact quantum
analysis of Sec.~\ref{sec:quantum}.

\subsection{From degeneracy point to exceptional point}
\label{subsec:quantum_confirmation}

\begin{figure}[htbp]
	\centering
	\includegraphics[height=5.5cm,width=7.50cm]{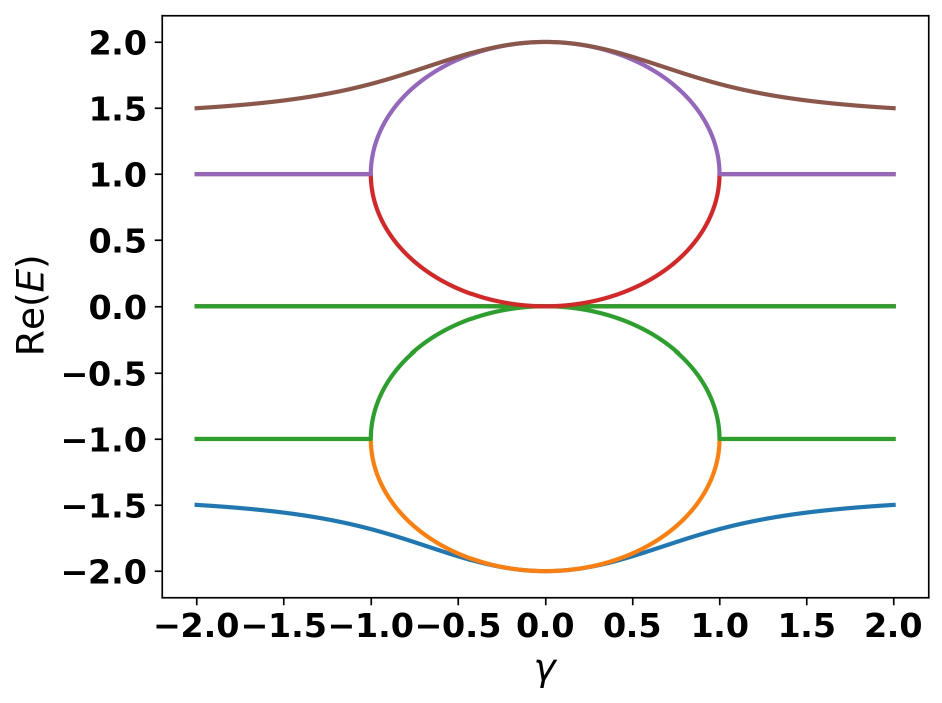}
	\includegraphics[height=5.5cm,width=7.50cm]{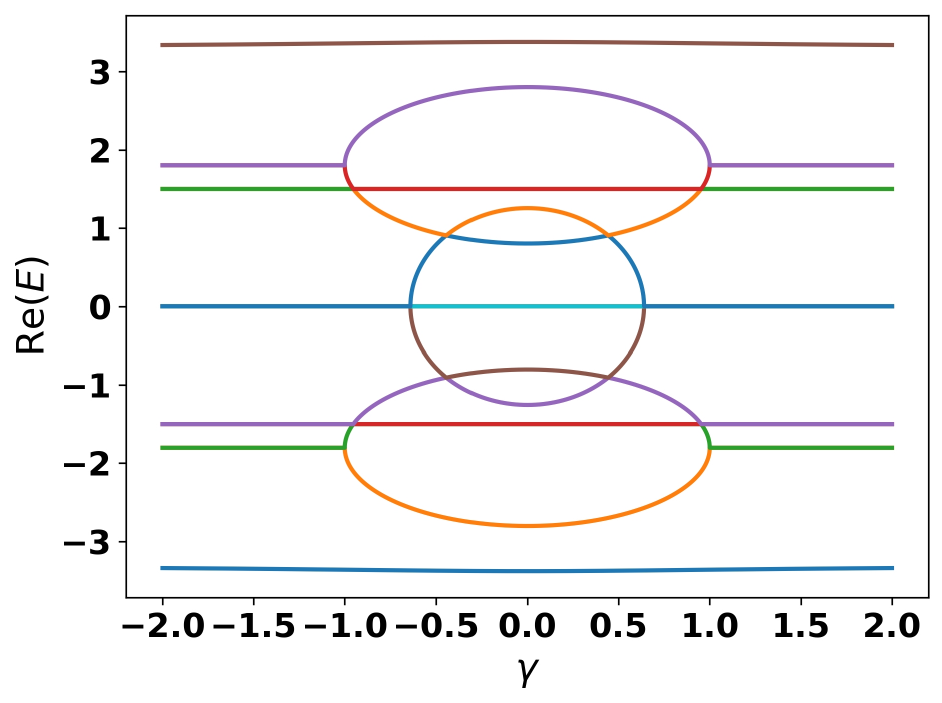}
    \includegraphics[height=5.5cm,width=7.50cm]{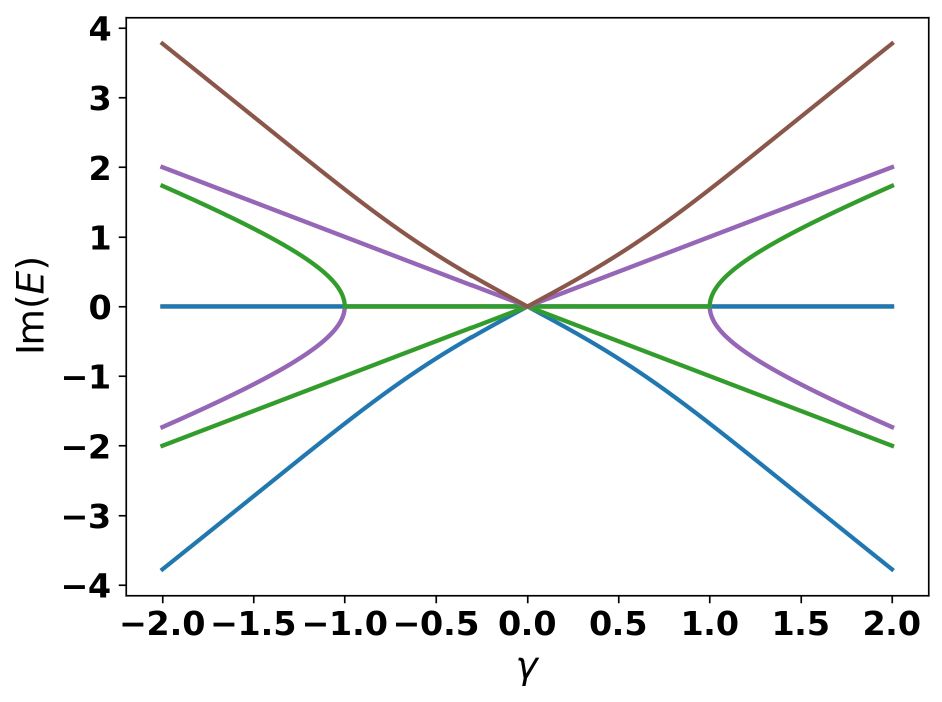}
	\includegraphics[height=5.5cm,width=7.50cm]{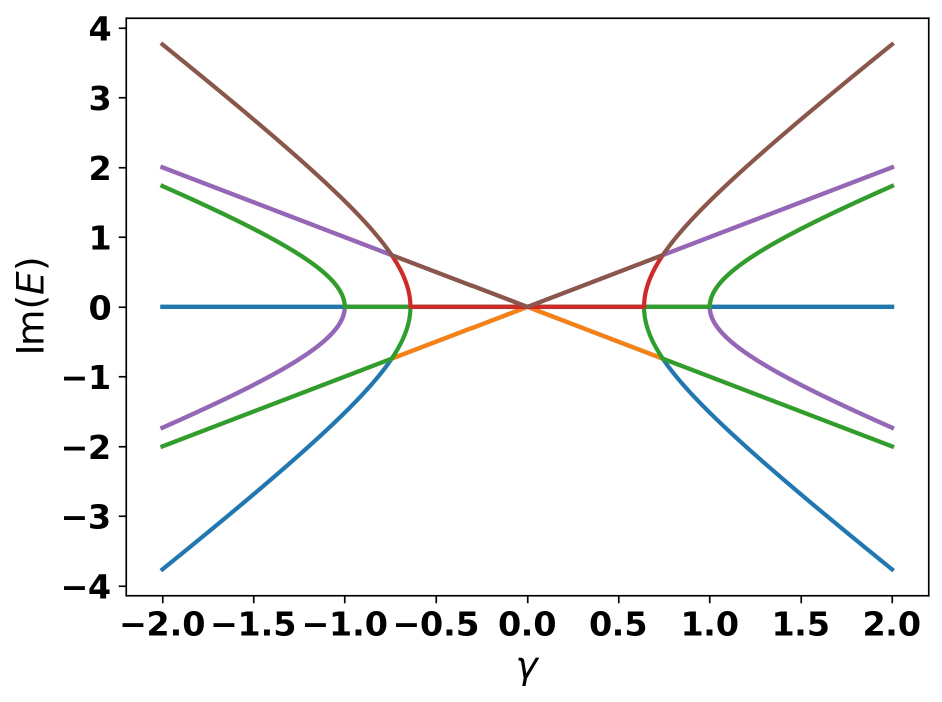}
	\caption{
		Finite-size quantum spectrum $N=4$. (a) Real part of the
		eigenenergies for $\eta/J=0.0$ . (b) Real part for $\eta/J=1.5$ (c) Imaginary part
		for $\eta/J=0.0$. (d) Imaginary part for $\eta/J=1.5$. In all cases, the point $|\gamma/J|=1$ is a point of spectral
		transition and this occurs independently of system size.
	}
	\label{fig:quantum_spectrum}
\end{figure}

To independently assess the critical structure identified from the classical
equilibrium analysis, we now examine the exact quantum spectrum obtained in
Sec.~\ref{sec:quantum}. Figure~\ref{fig:quantum_spectrum} displays the real
and imaginary parts of the many-body eigenenergies as functions of the
staggered non-Hermitian strength $\gamma/J$ for representative system
size $N=4$ at $J=1$,  $\eta/J=0.0$ and $\eta/J=1.5$.

For $\eta/J=0.0$, the real-energy spectrum shown in panel (a) forms a characteristic
closed branch structure within the interval $|\gamma|<1$. The energies of coalescing branches remain
real throughout this region and evolve continuously as $\gamma/J$ is varied
from negative to positive values. As $\gamma/J$ approaches either $-1$ or
$+1$, the separation between the relevant branches decreases, and the branches
meets. Beyond these points, the  real part of the corresponding
energies remain identical (separation collapses toward zero), while the corresponding imaginary parts,
shown in panel (c), become nonzero (complex conjugates). For $|\gamma/J|<1$, the imaginary parts
vanish, indicating that the relevant quantum energies remain entirely real.
At $|\gamma/J|=1$, the imaginary branches emerge from the degeneracy point,
and for $|\gamma|>1$ they develop equal and opposite imaginary components.
The transition at $|\gamma/J|=1$ therefore separates a real-energy regime
from a regime in which the quantum energies occur as complex-conjugate pairs. 
Moreover, we observe an accidental degeneracy involving external branches at $\gamma=0$; this similar 
behaviour was also observed in a classical analysis. at this point, several branches meets giving rise to a reach and complex landscape structure.

The $\eta/J=1.5$ results in panels (b) and (d) exhibit the same qualitative
structure but with more critical points(2). The real parts of the quantum energies again form branches that
converge at $\gamma=\pm1$, while the imaginary parts remain zero inside the
interval $|\gamma/J|<1$ and become nonzero outside it. The close
correspondence between the two spectra is significant because it
demonstrates that the critical scale is not an artefact of the system parameter choice. Interestingly, increasing the system size introduces additional many-body
states and therefore a richer spectral structure, but the characteristic
non-Hermitian threshold remains fixed at $|\gamma|=|J|$. Moreover, additional
degeneracy points emerge at momenta satisfying Eq.~\eqref{eq:k_EP}; these are
quantised according to $k_n=\pi n/M$ and lie within the interval $[0,\pi]$.
The degeneracies at the Brillouin-zone boundaries $k=0$ and $k=\pi$,
corresponding to $|\gamma|=|J|$, are particularly robust: they appear
irrespective of the parameter values $\eta$ and $J$ and independently of the
number of sites. This robustness is the finite-size fingerprint of the
exceptional-point supersurface identified analytically in
Sec.~\ref{sec:quantum}.

The vanishing of the energy separation at $\gamma=\pm J$ establishes spectral
degeneracy, but an exceptional point requires, in addition, eigenvector
coalescence and loss of diagonalizability. These latter properties were
established analytically from the BdG eigenvectors in Sec.~\ref{sec:quantum},
where the particle- and hole-sector eigenvectors were shown to become
linearly dependent at $\gamma=\pm J$, leaving the BdG matrix defective. Taken
together, the eigenvalue degeneracy and the eigenvector coalescence provide a
clear demonstration that the degeneracy point responsible for the degeneracy
of the equilibrium energies and for the merging of the classical state
configurations is an exceptional point of the non-Hermitian quantum system.

The quantum results therefore provide an independent confirmation of the
central classical finding. At the energy level, the classical equilibrium
energies and the quantum BdG spectrum become degenerate at exactly the same
family-independent critical scale $|\gamma|=|J|$, and both exhibit the
characteristic square-root branch-point structure. At the state level, the
classical configurations merge at the family-dependent values $\gamma=-J$ and
$\gamma=+J$, and the correspondence with the quantum description is through
the shared branch-point topology rather than through a direct comparison of
critical scales. Now that the exceptional character of the degeneracy point
has been established, it remains to verify its branch-point topology through
the dynamics of encirclement and the Riemann-sheet structure.

\subsection{Encirclement dynamics and branch-point topology}
\label{subsec:encirclement}

\begin{figure}[htbp]
	\centering
	 \includegraphics[width=17cm,height=11cm]{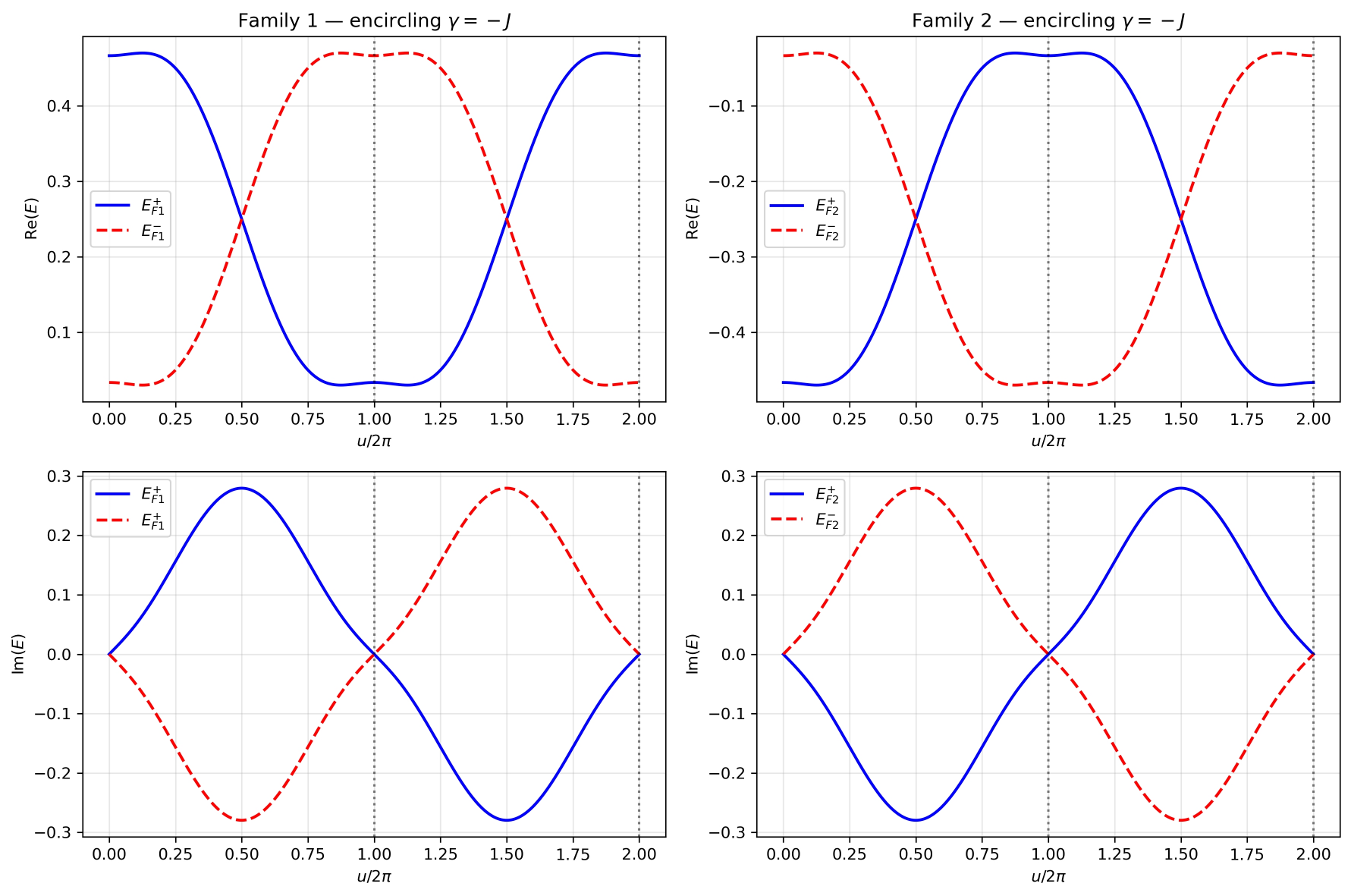}
	\caption{
		Encirclement dynamics of the energy branches around the degeneracy
		point at $\eta/J=0.1$. (a) Real part of the energy for Family~1.
		(b) Real part for Family~2. (c) Imaginary part for Family~1.
		(d) Imaginary part for Family~2. The two branches meet and exchange
		their identities after a single complete encirclement, confirming
		the branch-point topology of the degeneracy.
	}
	\label{fig:encirclement}
\end{figure}

To confirm the exceptional character of the degeneracy point, we examine
whether the energy branches possess the multivalued connectivity expected
when the degeneracy is encircled in the complex $\gamma$ plane. We define the
closed path
\begin{equation}
	\gamma(u)=\gamma_c+R\,e^{iu},
	\qquad
	0\leq u\leq 2\pi,
	\label{eq:encirclement_path}
\end{equation}
where $\gamma_c$ is the degeneracy point and $R$ is the radius of the path.
Figure~\ref{fig:encirclement} displays the energy landscape of Families~1
and~2 as the point $\gamma=J$ is encircled.

Figure~\ref{fig:encirclement} shows the evolution of the real and imaginary
parts of the energy branches as the complex control parameter traverses a
closed path centred at $\gamma=J$. For both families, the two branches meet,
exchange their positions during a single encirclement, and return to their
original identities only after a second complete circuit. The real and
imaginary components of the two branches vary continuously along the path,
but the branch trajectories do not return to their original identities after
one circuit. Instead, the two branches are analytically connected through the
degeneracy, so that a single encirclement maps one branch onto the other. A
second encirclement is required to return to the original branch. This
behaviour is the characteristic signature of a branch point and provides
direct confirmation of the exceptional-point character of the degeneracy at
$\gamma=J$.

The origin of this behaviour can be traced directly to the square-root factor
$S=\sqrt{J^2-\gamma^2}$. A single encirclement of a branch point changes the
sign of the square root, and since the two energy branches differ precisely
by this sign, one complete encirclement exchanges the two branches. This
two-sheet connectivity is stronger evidence of branch-point degeneracy than
the simple observation that two energies become equal on the real axis,
because it probes the global topology of the spectrum rather than a single
point.

The different locations of the Family~1 and Family~2 state-level merging are
again evident in Fig.~\ref{fig:riemann_sheet}. The two families exhibit
essentially the same branch topology, but the branch point is centred at
opposite values of $\gamma$. This reinforces the symmetry relation already
identified from the classical observables: the transformation
$\gamma\rightarrow-\gamma$ interchanges the locations of the two critical
points while preserving the qualitative branch geometry.

The encirclement result therefore establishes the global topology of the
energy degeneracy. The degeneracy is not merely a point at which two
numerical curves happen to intersect; it is a branch-point singularity
possessing non-trivial connectivity in the complex control-parameter plane.
This topology is the defining characteristic that distinguishes an
exceptional point from an ordinary degeneracy, and its identification
completes the demonstration that the classical observable coalescence
observed in Sec.~\ref{subsec:classical_fusion} is the physical manifestation
of a quantum exceptional point.

By contrast, the additional degeneracy at $\gamma=0$ behaves fundamentally
differently. As established in Fig.~\ref{fig:DE}(b), the inter-family energy
separation vanishes at this point, indicating a degeneracy of the equilibrium
energies. However, the inter-family observable separation $D_X$, shown in
Fig.~\ref{fig:DX}(b), does not vanish at $\gamma=0$: the two classical
configurations belonging to different families remain distinguishable in the
physical observable space even though their energies coincide. This is the
first indication that the degeneracy at $\gamma=0$ is not an exceptional
point. The second indication is the encirclement behaviour: a closed path
around $\gamma=0$ does not exchange the branches, in contrast with the
exchange observed at $\gamma=J$. The combined evidence from the energy
separation, the observable separation, and the encirclement topology
identifies the degeneracy at $\gamma=0$ as an ordinary degeneracy of the kind
familiar from Hermitian physics: a diabolic point. It is not accompanied by
state coalescence, defectiveness of the underlying operator, or non-trivial
branch-point topology, and it therefore stands in sharp contrast to the
exceptional point at $|\gamma|=|J|$.

\subsection{Riemann-sheet structure of the exceptional point}
\label{subsec:riemann}

\begin{figure}[htbp]
	\centering
	\includegraphics[width=8cm,height=8cm]{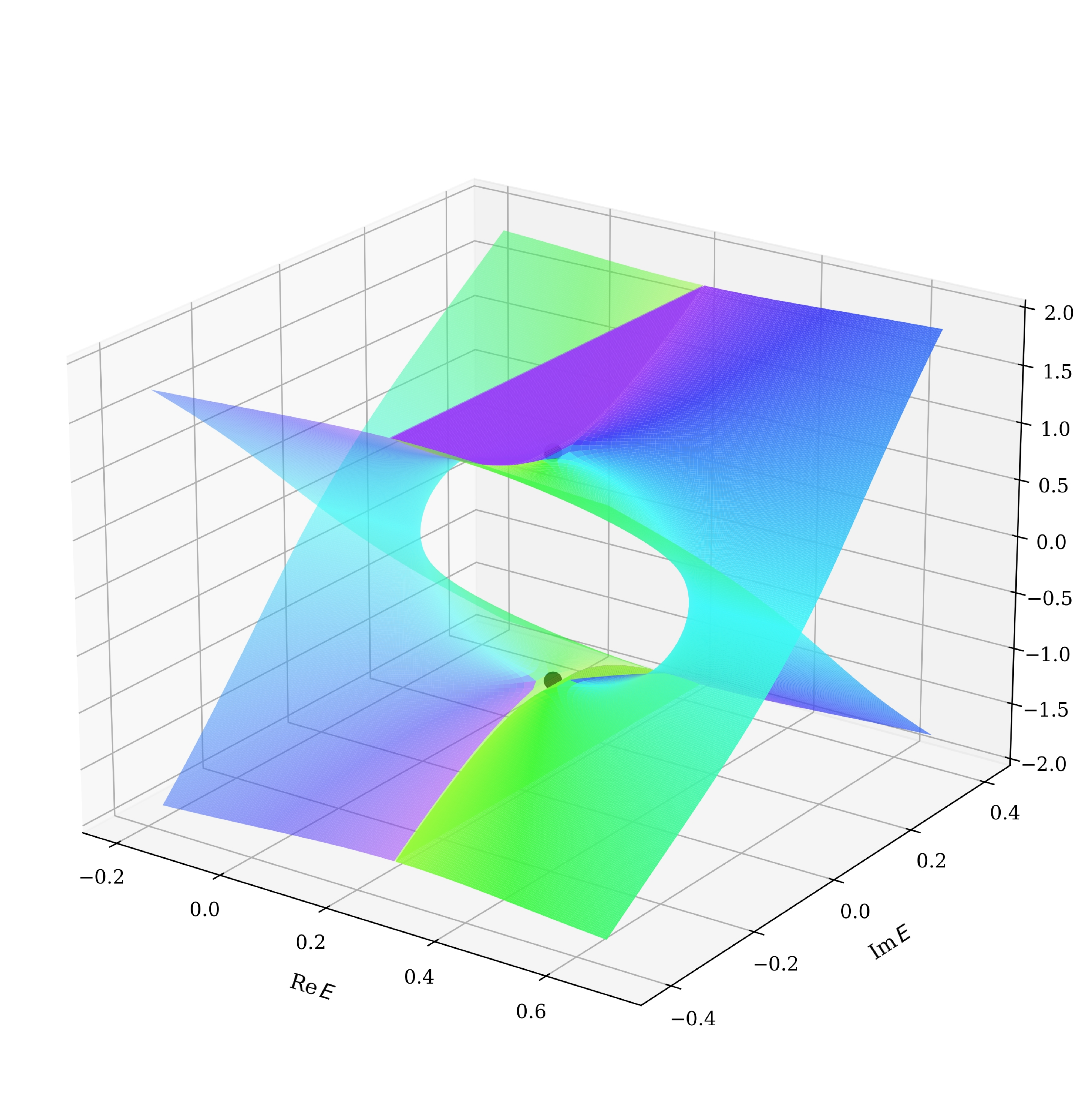}
	\includegraphics[width=8cm,height=8cm]{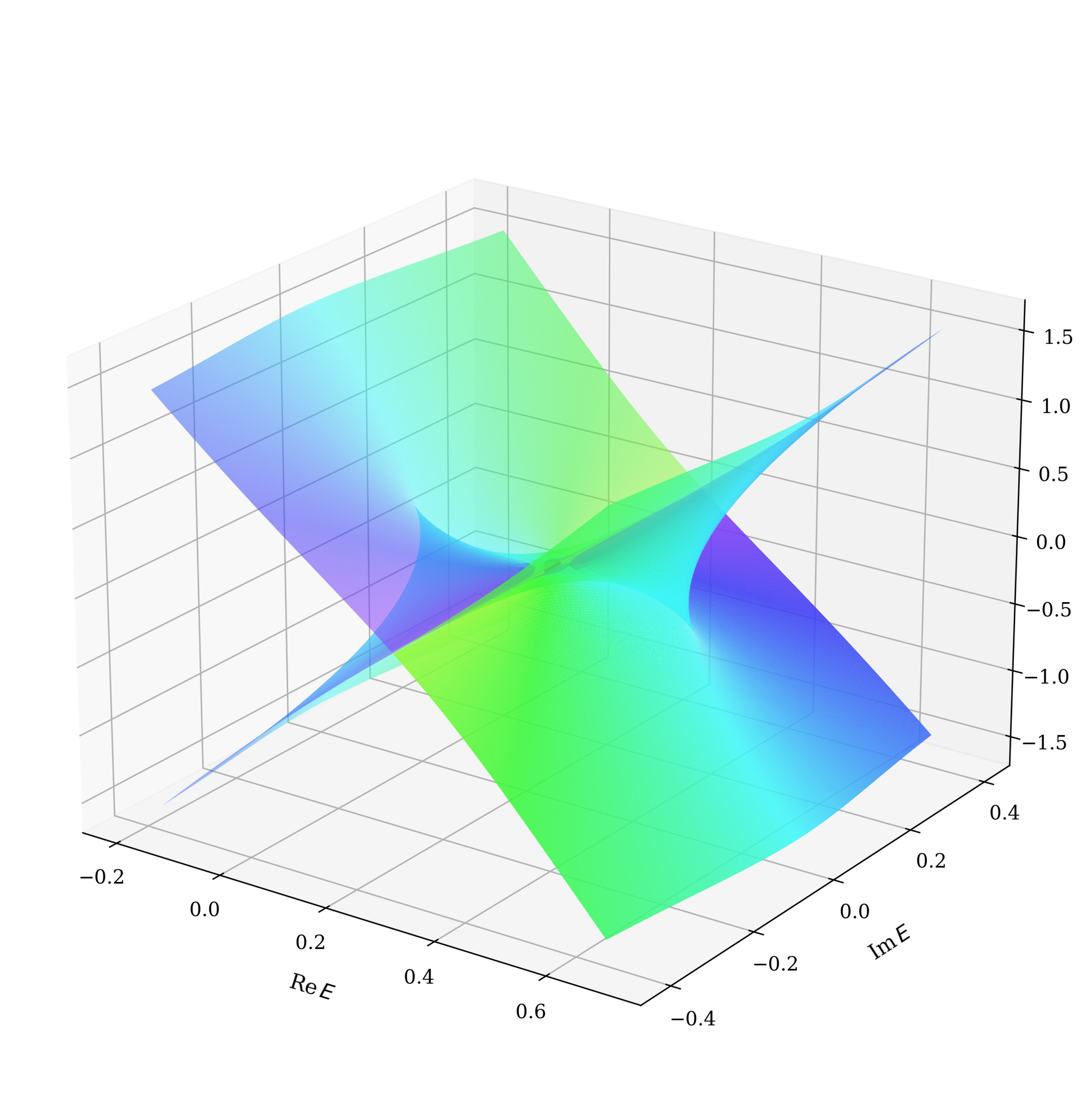}
	\caption{
		Riemann sheets of the classical equilibrium energies around the
		exceptional point at $\gamma=J$, displayed in the complex $\gamma$
		plane. (a) real part of the energy. (b) imaginary part of the
		energy. The two sheets meet at the branch point and are analytically
		connected through it, so that a continuous path encircling the
		branch point transfers the solution from one sheet to the other.
	}
	\label{fig:riemann_sheet}
\end{figure}

A further element that confirms the exceptional-point nature of the
degeneracy is the Riemann-sheet structure of the classical energy branches.
We plot the Riemann sheets of the equilibrium energies for the two families
around the branch point in Fig.~\ref{fig:riemann_sheet}.

Figure~\ref{fig:riemann_sheet} plots the Riemann-sheet structure obtained for
the known exceptional point at $\gamma=J$. The two surfaces displayed for the
real and imaginary parts of the energy illustrate a two-sheeted structure.
Each family possesses two branches, which meet at the corresponding branch
point. For $\gamma=J$, the two sheets are not independent surfaces; they are
analytically connected through the branch point, and the apparent merging of
the branches on the real axis is the visible trace of this global
Riemann-sheet structure. The geometry of the sheets explains how the two
classical equilibrium configurations connect across the parameter plane: a
continuous path that encircles the branch point transfers the solution from
one sheet to the other, so that the branch identity is path-dependent. This
is precisely the topology revealed independently by the encirclement
calculation of Fig.~\ref{fig:encirclement}, and it is entirely absent at the
diabolic point discussed at the end of Sec.~\ref{subsec:encirclement}.

The Riemann-sheet construction is thus presented as an analytic continuation
of the equilibrium branch quantities and is directly relevant to the
classical branch analysis. The combined results demonstrate that, in the
non-Hermitian system considered here, classical configurations coalesce at an
exceptional point located on the boundary of the exceptional-point
supersurface, while the additional degeneracy at $\gamma=0$ corresponds to a
diabolic point and is therefore not accompanied by exceptional-point
topology. This distinction supports the principal objective of the study:
classical observable coalescence provides a physically meaningful
manifestation of a non-Hermitian degeneracy, and it is capable of
discriminating between different types of spectral singularity through the
combination of observable separation, energy separation, encirclement
behaviour, and Riemann-sheet connectivity.

\subsection{Synthesis and discussion}
\label{subsec:synthesis}

The results presented above establish a multi-level correspondence between
the classical and quantum descriptions of the non-Hermitian spin chain. At
the energy level, the correspondence is exact. The classical equilibrium
energies exhibit the square-root dependence $S=\sqrt{J^2-\gamma^2}$ and
become degenerate at the family-independent critical scale $|\gamma|=|J|$;
the exact BdG spectrum exhibits the same square-root branch-point singularity
and the same critical scale. The vanishing of the classical energy separation
$D_E$ and the quantum spectral separation occurs at precisely the same value
of $\gamma$, and this agreement is the more significant because it holds
independently of system size and of the real transverse field $\eta$. At the
state level, the classical equilibrium configurations merge at the
family-dependent values $\gamma=-J$ (Family~1) and $\gamma=+J$ (Family~2).
This merging is not directly comparable with the quantum spectrum, which is a
statement about eigenenergies. Its connection to the exceptional point is
established instead through the shared branch-point topology: the same
square-root quantity $S$ that controls the energy-level degeneracy also
controls the separation of the classical state configurations, and the
analytic continuation of the classical branches into the complex $\gamma$
plane reveals a two-sheeted Riemann structure whose branch points coincide
with the exceptional points of the quantum spectrum. The state-level merging
is therefore the observable manifestation of the same underlying
non-Hermitian critical structure, confirmed by the encirclement and
Riemann-sheet analyses.

The distinction between the exceptional point at $|\gamma|=|J|$ and the
diabolic point at $\gamma=0$ is the discriminating result of this work. Both
appear as degeneracies in the energy spectrum, but only the former is
accompanied by a genuine loss of distinguishability of the classical
configurations. At $\gamma=0$, the inter-family energy separation $D_E$
vanishes, but the inter-family observable separation $D_X$ remains finite,
and the encirclement of this point does not exchange the branches. The
exceptional point at $|\gamma|=|J|$ exhibits eigenvector coalescence,
defectiveness of the quantum operator, branch exchange under encirclement,
and connected Riemann sheets; the diabolic point at $\gamma=0$ exhibits none
of these features. The ability to distinguish these two cases within a single
model illustrates the importance of the multi-level diagnostic approach
adopted here. Energy degeneracy alone is insufficient to identify an
exceptional point; it must be accompanied by observable coalescence and by
the topological signature of branch exchange under encirclement, or,
equivalently, by eigenvector coalescence in the exact quantum spectrum.
Conversely, the observation of an energy degeneracy that is not accompanied
by observable coalescence and does not produce branch exchange under
encirclement is a positive identification of a diabolic point, which is
itself a useful diagnostic.

The results have implications beyond the specific model considered here. The
notion of classical observable coalescence developed in this work is
applicable to any nonlinear classical system whose linearised dynamics is
governed by a non-Hermitian operator. Since effective non-Hermitian operators
arise naturally in dissipative and driven classical systems, including
coupled electrical circuits, mechanical oscillators, optical resonators,
acoustic networks, and Josephson-junction arrays \cite{heiss2012physics,
moiseyev2011non,schindler2011experimental,coulais2017static,
fernandez2026parity}, the observable-level description proposed here can in
principle be extended to these platforms. 

Moreover, the semiclassical analysis of this work is based on a mean-field reduction of
the Heisenberg equations of motion combined with the spin-coherent-state
(Bloch-sphere) representation of the local two-level systems. This choice of
framework is not a limitation but rather the appropriate methodology for the
present study. The objective is to extract the classical observables of the
non-Hermitian spin chain, and the spin-coherent-state representation places
the analysis exactly at the quantum-classical boundary where these
observables are defined. In this regime, collective behaviour dominates over
individual quantum dynamics, and higher-order correlations do not modify the
equilibrium structure; the classical limit is not an approximation to be
improved upon but the correct description of the observables of interest. The
Bloch-sphere parametrisation further accommodates the non-Hermiticity of the
model through the global phase $\lambda_\nu$, which absorbs the
non-Hermitian contribution and ensures that the local state remains
normalizable even when the Hamiltonian is non-Hermitian. The agreement
between the semiclassical equilibrium energies and the exact quantum
spectrum, established in Sec.~\ref{subsec:quantum_confirmation}, provides
independent confirmation that the classical critical structure is faithful to
the underlying quantum problem. The analysis has been restricted to the
specific case of a two-sublattice Ising-like chain with a staggered imaginary
field; extensions to other non-Hermitian spin models, to higher dimensions,
and to systems with long-range interactions would be natural directions for
future work.

Notwithstanding this scope, the central result of this work is robust:
classical observable coalescence and quantum exceptional-point physics are
distinct yet intimately related manifestations of a common non-Hermitian
critical structure. The energy-level correspondence is exact; the state-level
merging is the observable manifestation of the same branch-point structure;
and the combination of observable separation, energy separation, encirclement
behaviour, and Riemann-sheet connectivity provides a discriminating
diagnostic capable of distinguishing exceptional points from diabolic points.
This framework establishes a conceptual bridge between nonlinear classical
equilibrium geometry and non-Hermitian spectral theory, and it suggests that
classical observables may serve as practical signatures of exceptional-point
physics in a wide range of dissipative and driven classical many-body
systems.

\section{Summary and Conclusions}
\label{sec:conclusions}

In this work we have investigated the coalescence of classical observables in
a non-Hermitian two-sublattice Ising-like system and established its
relationship with the exceptional-point structure of the corresponding
quantum model. The classical configurations were obtained from a mean-field
treatment of the Heisenberg dynamics within a spin-coherent-state
representation, which yields the equilibrium branches and their energies,
while the quantum spectrum was obtained exactly through Jordan-Wigner and
Bogoliubov-de Gennes diagonalisation.

The classical equilibrium configurations organise into two symmetry-related
families, each containing two branches that lose their distinguishability at
the family-dependent values $\gamma=-J$ (Family~1) and $\gamma=+J$
(Family~2). The equilibrium energies become degenerate at the
family-independent scale $|\gamma|=|J|$, and analytic continuation into the
complex $\gamma$ plane reveals a two-sheeted Riemann structure whose branch
points coincide with the critical values. The exact quantum spectrum confirms
the exceptional character of this degeneracy, and encirclement in the complex
$\gamma$ plane exchanges both the energy branches and the fixed points,
providing a direct topological signature of the branch-point structure. A
comparison with the additional degeneracy at $\gamma=0$ identifies the latter
as a diabolic point rather than an exceptional point.

These results show that the non-Hermitian parameter $\gamma$ provides direct
control over the distinguishability of the classical equilibrium
configurations, and that the energy-level critical scale is independently
reproduced by the quantum theory. Classical observable coalescence and
quantum exceptional-point physics therefore emerge as distinct yet rigorously
related manifestations of a common non-Hermitian critical structure. The
multi-level diagnostic combining observable separation, energy separation,
Riemann-sheet connectivity, and eigenvector coalescence provides a
discriminating test between exceptional and diabolic points, and the
observable-level description offers a physically transparent, experimentally
accessible perspective on non-Hermitian degeneracies in dissipative and
driven classical many-body systems. Extensions to other non-Hermitian spin
models, to higher-dimensional systems, and to experimental platforms where
classical observables can be directly measured constitute natural directions
for future work.

\bibliography{references}

@article{bender1998real,
  author  = {Bender, Carl M. and Boettcher, Stefan},
  title   = {Real Spectra in Non-Hermitian Hamiltonians Having
             $\mathcal{PT}$ Symmetry},
  journal = {Physical Review Letters},
  volume  = {80},
  number  = {24},
  pages   = {5243--5246},
  year    = {1998},
  doi     = {10.1103/PhysRevLett.80.5243}
}

@article{ashida2020non,
  author  = {Ashida, Yuto and Gong, Zongping and Ueda, Masahito},
  title   = {Non-Hermitian Physics},
  journal = {Advances in Physics},
  volume  = {69},
  number  = {3},
  pages   = {249--435},
  year    = {2020},
  doi     = {10.1080/00018732.2021.1876991}
}

@article{el2007theory,
  author  = {El-Ganainy, Ramy and Makris, Konstantinos G. and
             Christodoulides, Demetrios N. and Musslimani, Ziad H.},
  title   = {Theory of Coupled Optical $\mathcal{PT}$-Symmetric Structures},
  journal = {Optics Letters},
  volume  = {32},
  number  = {17},
  pages   = {2632--2634},
  year    = {2007},
  doi     = {10.1364/OL.32.002632}
}

@article{heiss2012physics,
  author  = {Heiss, W. D.},
  title   = {The Physics of Exceptional Points},
  journal = {Journal of Physics A: Mathematical and Theoretical},
  volume  = {45},
  number  = {44},
  pages   = {444016},
  year    = {2012},
  doi     = {10.1088/1751-8113/45/44/444016}
}

@book{moiseyev2011non,
  author    = {Moiseyev, Nimrod},
  title     = {Non-Hermitian Quantum Mechanics},
  publisher = {Cambridge University Press},
  address   = {Cambridge},
  year      = {2011},
  isbn      = {9780521889728},
  doi       = {10.1017/CBO9780511976186}
}

@book{kato1966perturbation,
  author    = {Kato, Tosio},
  title     = {Perturbation Theory for Linear Operators},
  publisher = {Springer},
  address   = {Berlin, Heidelberg},
  year      = {1966},
  doi       = {10.1007/978-3-662-12678-3}
}

@article{rotter2009non,
  author  = {Rotter, Ingrid},
  title   = {A Non-Hermitian Hamilton Operator and the Physics
             of Open Quantum Systems},
  journal = {Journal of Physics A: Mathematical and Theoretical},
  volume  = {42},
  number  = {15},
  pages   = {153001},
  year    = {2009},
  doi     = {10.1088/1751-8113/42/15/153001}
}

@article{berry2004physics,
  author  = {Berry, M. V.},
  title   = {Physics of Nonhermitian Degeneracies},
  journal = {Czechoslovak Journal of Physics},
  volume  = {54},
  number  = {10},
  pages   = {1039--1047},
  year    = {2004},
  doi     = {10.1023/B:CJOP.0000044002.05657.04}
}

@article{miri2019exceptional,
  author  = {Miri, Mohammad-Ali and Al{\`u}, Andrea},
  title   = {Exceptional Points in Optics and Photonics},
  journal = {Science},
  volume  = {363},
  number  = {6422},
  pages   = {eaar7709},
  year    = {2019},
  doi     = {10.1126/science.aar7709}
}

@article{ryu2024exceptional,
  author  = {Ryu, Jung-Wan and Han, Jae-Ho and Yi, Chang-Hwan and
             Park, Moon Jip and Park, Hee Chul},
  title   = {Exceptional Classifications of Non-Hermitian Systems},
  journal = {Communications Physics},
  volume  = {7},
  number  = {1},
  pages   = {109},
  year    = {2024},
  doi     = {10.1038/s42005-024-01595-9}
}

@article{ozdemir2019parity,
  author  = {{\"O}zdemir, {\c{S}}ahin K. and Rotter, Stefan and
             Nori, Franco and Yang, Lan},
  title   = {Parity--Time Symmetry and Exceptional Points in Photonics},
  journal = {Nature Materials},
  volume  = {18},
  number  = {8},
  pages   = {783--798},
  year    = {2019},
  doi     = {10.1038/s41563-019-0304-9}
}

@article{peng2014parity,
  author  = {Peng, Bo and {\"O}zdemir, {\c{S}}ahin K. and Lei, Fuchuan and
             Monifi, Faraz and Gianfreda, Mariagiovanna and Long, Gui Lu and
             Fan, Shanhui and Nori, Franco and Bender, Carl M. and Yang, Lan},
  title   = {Parity--Time-Symmetric Whispering-Gallery Microcavities},
  journal = {Nature Physics},
  volume  = {10},
  number  = {5},
  pages   = {394--398},
  year    = {2014},
  doi     = {10.1038/nphys2927}
}

@article{feng2013experimental,
  author  = {Feng, Liang and Xu, Ye-Long and Fegadolli, William S. and
             Lu, Ming-Hui and Oliveira, Jos{\'e} E. B. and
             Almeida, Vilson R. and Chen, Ye-Feng and Scherer, Axel},
  title   = {Experimental Demonstration of a Unidirectional
             Reflectionless Parity--Time Metamaterial at Optical Frequencies},
  journal = {Nature Materials},
  volume  = {12},
  number  = {2},
  pages   = {108--113},
  year    = {2013},
  doi     = {10.1038/nmat3495}
}

@article{schindler2011experimental,
  author  = {Schindler, Joseph and Li, Ang and Zheng, Mei C. and
             Ellis, F. M. and Kottos, Tsampikos},
  title   = {Experimental Study of Active {LRC} Circuits with
             $\mathcal{PT}$ Symmetries},
  journal = {Physical Review A},
  volume  = {84},
  number  = {4},
  pages   = {040101},
  year    = {2011},
  doi     = {10.1103/PhysRevA.84.040101}
}

@article{coulais2017static,
  author  = {Coulais, Corentin and Sounas, Dimitrios and Al{\`u}, Andrea},
  title   = {Static Non-Reciprocity in Mechanical Metamaterials},
  journal = {Nature},
  volume  = {542},
  number  = {7642},
  pages   = {461--464},
  year    = {2017},
  doi     = {10.1038/nature21044}
}

@article{fernandez2026parity,
  author  = {Fern{\'a}ndez-Alc{\'a}zar, Lucas J. and Zhong, Qi and
             Kuhl, Ulrich and Chen, Pai-Yen and El-Ganainy, Ramy and
             Kottos, Tsampikos},
  title   = {Parity--Time Symmetry and Exceptional Points in
             Electronic Circuits},
  journal = {Nature Electronics},
  volume  = {9},
  number  = {5},
  year    = {2026},
  doi     = {10.1038/s41928-026-00542-7}
}

@article{wiersig2014enhancing,
  author  = {Wiersig, Jan},
  title   = {Enhancing the Sensitivity of Frequency and Energy
             Splitting Detection by Using Exceptional Points:
             Application to Microcavity Sensors for Single-Particle Detection},
  journal = {Physical Review Letters},
  volume  = {112},
  number  = {20},
  pages   = {203901},
  year    = {2014},
  doi     = {10.1103/PhysRevLett.112.203901}
}

@article{castro2009spin,
  author  = {Castro-Alvaredo, Olalla A. and Fring, Andreas},
  title   = {A Spin Chain Model with Non-Hermitian Interaction:
             The Ising Quantum Spin Chain in an Imaginary Field},
  journal = {Journal of Physics A: Mathematical and Theoretical},
  volume  = {42},
  number  = {46},
  pages   = {465211},
  year    = {2009},
  doi     = {10.1088/1751-8113/42/46/465211}
}

@article{starkov2023quantum,
  author  = {Starkov, Grigory A. and Fistul, Mikhail V. and
             Eremin, Ilya M.},
  title   = {Quantum Phase Transitions in Non-Hermitian
             $\mathcal{PT}$-Symmetric Transverse-Field Ising Spin Chains},
  journal = {Annals of Physics},
  volume  = {456},
  pages   = {169268},
  year    = {2023},
  doi     = {10.1016/j.aop.2023.169268}
}

@article{lenke2021high,
  author  = {Lenke, Lea and M{\"u}hlhauser, Matthias and
             Schmidt, Kai Phillip},
  title   = {High-Order Series Expansion of Non-Hermitian Quantum Spin Models},
  journal = {Physical Review B},
  volume  = {104},
  number  = {19},
  pages   = {195137},
  year    = {2021},
  doi     = {10.1103/PhysRevB.104.195137}
}

@article{muldoon2023pt,
  author  = {Muldoon, Jacob and Joglekar, Yogesh N.},
  title   = {$\mathcal{PT}$-Symmetry Breaking in Quantum Spin Chains
             with Exceptional Non-Hermiticities},
  journal = {Physical Review A},
  volume  = {108},
  number  = {6},
  pages   = {062205},
  year    = {2023},
  doi     = {10.1103/PhysRevA.108.062205}
}

@article{meisinger2012pt,
  author  = {Meisinger, Peter N. and Ogilvie, Michael C.},
  title   = {$\mathcal{PT}$ Symmetry in Classical and Quantum
             Statistical Mechanics},
  journal = {Philosophical Transactions of the Royal Society A},
  volume  = {371},
  number  = {1989},
  pages   = {20120058},
  year    = {2013},
  doi     = {10.1098/rsta.2012.0058}
}

@article{Maffi2024,
  title = {Quantum Simulation of the Tricritical Ising Model in Tunable Josephson Junction Ladders},
  author = {Maffi, Lorenzo and Tausendpfund, Niklas and Rizzi, Matteo and Burrello, Michele},
  journal = {Physical Review Letters},
  volume = {132},
  issue = {22},
  pages = {226502},
  year = {2024},
  doi = {10.1103/PhysRevLett.132.226502},
  note = {Hybrid superconductor-semiconductor Josephson junction array in a two-leg ladder geometry for simulating the tricritical Ising phase transition}
}

@article{Razmkhah2023,
  title = {A Josephson Parametric Oscillator-Based Ising Machine},
  author = {Razmkhah, Sasan and others},
  journal = {Physical Review B},
  volume = {109},
  pages = {014512},
  year = {2024},
  doi = {10.1103/PhysRevB.109.014512},
  eprint = {2309.03407},
  archivePrefix = {arXiv},
  note = {JPO-based tile structure operating at 7.5 GHz; compatible with Lechner-Hauke-Zoller architecture for scalable superconductor-based Ising machines}
}

@inproceedings{Belkhadir2025,
  title     = {An Ising Machine Based on Millimeter-Wave Coupled LC Oscillators},
  author    = {Belkhadir, A. and Banerjee, A.},
  booktitle = {2025 IEEE International Symposium on Circuits and Systems (ISCAS)},
  year      = {2025},
  address   = {London, United Kingdom},
  doi       = {10.1109/ISCAS56056.2025.11043357},
  note      = {45nm CMOS technology, oscillators at 32.2 GHz; solves Max-Cut within 0.5 ns using second harmonic injection at tail transistor gates and tank injection coupling}
}

@article{McGoldrick2022,
  title = {Ising Machine Based on Electrically Coupled Spin Hall Nano-Oscillators},
  author = {McGoldrick, Brooke C. and Sun, Jonathan Z. and Liu, Luqiao},
  journal = {Physical Review Applied},
  volume = {17},
  issue = {1},
  pages = {014006},
  year = {2022},
  doi = {10.1103/PhysRevApplied.17.014006},
  note = {Dipole-coupled SHNOs modeled via coupled Landau-Lifshitz-Gilbert-Slonczewski equations; electrically coupled SHNO Ising machine}
}

@article{Cowburn2023,
  title = {Electrically programmable magnetic coupling in an Ising network exploiting solid-state ionic gating},
  author = {Cowburn, Russell P. and others},
  journal = {Nature Communications},
  volume = {14},
  pages = {6393},
  year = {2023},
  doi = {10.1038/s41467-023-42126-8},
  note = {Coupled nanomagnets with voltage-controlled phase transition in artificial spin ices; magnetization states act as Ising spins with tunable coupling}
}

@article{Iftakher2026,
  title = {Intrinsic annealing in a hybrid memristor-magnetic tunnel junction Ising machine},
  author = {Iftakher, I. and others},
  journal = {Nature Communications},
  year = {2026},
  note = {Combines hafnium-oxide memristor crossbar arrays (storing multi-level couplings) with stochastic magnetic tunnel junctions (SMTJs) acting as thermally driven spins; the same read voltage enables intrinsic voltage-controlled annealing}
}

@article{Roushan2017,
  title = {Observing Parity-Time Symmetry Breaking in a Josephson Parametric Amplifier},
  author = {Roushan, P. and others},
  journal = {Physical Review Research},
  volume = {5},
  pages = {L042024},
  year = {2023},
  doi = {10.1103/PhysRevResearch.5.L042024},
  note = {Degenerate parametric amplifier in three-wave mixing mode; realizes balanced gain and loss between the two quadrature modes of the amplifier for PT-symmetry breaking}
}

@article{Zhang2025,
  title = {Topological Eigenvalue Braiding and Quantum State Transfer Near a Third-Order Exceptional Point},
  author = {Zhang, H. and Liu, T. and Xiang, Z. and others},
  journal = {PRX Quantum},
  volume = {6},
  issue = {2},
  pages = {020328},
  year = {2025},
  doi = {10.1103/PRXQuantum.6.020328},
  note = {Experimental investigation of eigenvalue braiding and state transfer arising from the encirclement of exceptional points in a three-dimensional non-Hermitian quantum system using superconducting circuits; demonstrates quasistatic encirclement with three independent parameters}
}

@article{Naghiloo2019,
  title = {Quantum state tomography across the exceptional point in a single dissipative qubit},
  author = {Naghiloo, M. and Abbasi, M. and Joglekar, Yogesh N. and Murch, K. W.},
  journal = {Nature Physics},
  volume = {15},
  issue = {12},
  pages = {1232--1236},
  year = {2019},
  doi = {10.1038/s41567-019-0652-z},
  note = {Post-selection on a three-level superconducting transmon circuit with tunable Rabi drive, dissipation, and detuning; quantum state tomography across the exceptional point reveals PT-symmetry breaking transition and decoherence enhancement}
}

@article{Schwinger1965,
  title = {On Angular Momentum},
  author = {Schwinger, Julian},
  journal = {Quantum Theory of Angular Momentum},
  editor = {Biedenharn, L. C. and Van Dam, H.},
  publisher = {Academic Press, New York},
  pages = {229--279},
  year = {1965},
  note = {Original formulation of the Schwinger boson representation of angular momentum}
}

@article{Auerbach1990,
  title = {New approaches to the quantum Heisenberg models: Schwinger boson representations},
  author = {Auerbach, Assa and Arovas, Daniel P.},
  journal = {Journal of Applied Physics},
  volume = {67},
  number = {9},
  pages = {5734--5736},
  year = {1990},
  doi = {10.1063/1.345947},
  note = {Invited review of the Schwinger boson representation and large-\(N\) mean-field theory for quantum Heisenberg models}
}

@article{Auerbach1994,
  title = {Bosonic mean-field theory of quantum Heisenberg spin systems: Bose condensation and magnetic order},
  author = {Auerbach, Assa and Arovas, Daniel P.},
  journal = {Physical Review B},
  volume = {38},
  number = {1},
  pages = {316--329},
  year = {1988},
  doi = {10.1103/PhysRevB.38.316},
  note = {Mean-field decoupling of spin operators in the Schwinger boson representation; includes the factorisation scheme used in this appendix}
}

@article{Radcliffe1971,
  title = {Some properties of coherent spin states},
  author = {Radcliffe, J. M.},
  journal = {Journal of Physics A: General Physics},
  volume = {4},
  number = {3},
  pages = {313--323},
  year = {1971},
  doi = {10.1088/0305-4470/4/3/009},
  note = {Original construction of spin coherent states (Bloch coherent states) parameterised by \((\theta,\phi)\) on the Bloch sphere}
}

@article{Gilmore1972,
  title = {Geometry of symmetrized states},
  author = {Gilmore, Robert},
  journal = {Annals of Physics},
  volume = {74},
  number = {2},
  pages = {391--463},
  year = {1972},
  doi = {10.1016/0003-4916(72)90147-9},
  note = {Systematic treatment of spin coherent states and their geometric interpretation on the Bloch sphere}
}

@article{Znojil2015,
  title = {Non-Hermitian Heisenberg representation},
  author = {Znojil, Miloslav},
  journal = {Physics Letters A},
  volume = {379},
  number = {36},
  pages = {2013--2017},
  year = {2015},
  doi = {10.1016/j.physleta.2015.06.024},
  note = {Formulation of the Heisenberg equation of motion for non-Hermitian Hamiltonians; the adjoint evolution is taken into account}
}

@article{Sans2026QuantumConversion,
  author  = {Sans, F. M. and Afungchui, D. and Vubangsi, M. and Tchoffo, M. and Fai, L. C. and Jipdi, M. N.},
  title   = {Modeling opinion dynamics in complex networks via quantum-theoretic frameworks: A quantum conversion approach},
  journal = {Chaos, Solitons \& Fractals},
  volume  = {203},
  pages   = {117650},
  year    = {2026},
  month   = feb,
  note    = {Article number 117650}
}

@article{Jipdi2026QuantumInterference,
  author  = {Jipdi, N. M. and Sans, F. M. and Afunchue, D. and Ateuafack, M. A. and Fai, L. C. and Tchoffo, M.},
  title   = {Quantum interference as a non-coercive mechanism for social consensus and collective memory formation},
  journal = {The European Physical Journal Plus},
  volume  = {141},
  number  = {3},
  pages   = {349},
  year    = {2026},
  month   = mar,
  note    = {Article number 349}
}

@misc{Jipdi2026TopologicallyProtected,
  author        = {Jipdi, M. N. and Mba, C. A. and Moubissi, A. B. and Fai, L. C. and Ateuafack, M. E.},
  title         = {Topologically Protected Learning from Exceptional Point Braiding: Toward Braid Programming},
  year          = {2026},
  month         = aug,
  eprint        = {2608.15829},
  archivePrefix = {arXiv},
  note          = {arXiv preprint arXiv:2608.15829}
}

@article{Sans2026QuantumHamiltonian,
  author  = {Sans, M. and Afunchue, M. and Ateuafack, M. A. and Tchoffo, M. and Fai, L. C. and Jipdi, M. N.},
  title   = {A Quantum Hamiltonian Framework for emergent phase transitions in opinion dynamics: Cognitive vs Somatic Interaction},
  journal = {Physics Letters A},
  pages   = {131921},
  year    = {2026},
  month   = jun,
  note    = {Article number 131921}
}

@article{wah2025manybody,
  title={Many-body neural network wavefunction for a non-Hermitian Ising chain},
  author={Wah, Lavoisier and Zen, Remmy and Kunst, Flore K.},
  journal={Physical Review Research},
  volume={7},
  pages={043291},
  year={2025},
  publisher={American Physical Society},
  doi={10.1103/q8p7-k7ms},
  url={https://journals.aps.org/prresearch/abstract/10.1103/q8p7-k7ms}
}

@article{wah2026bridging,
  title={Bridging Frustration and Non-Hermiticity via COMPASS: An Adaptive Biorthogonal Neural Quantum State Framework},
  author={Wah, Lavoisier and Kunst, Flore K and Hibat-Allah, Mohamed},
  journal={arXiv preprint arXiv:2607.13790},
  year={2026},
  url={https://arxiv.org/abs/2607.13790}
}

\appendix
\section{Experimental Platforms for the Non-Hermitian Spin Chain}
\label{app:platforms}

The non-Hermitian spin chain described by Eq.~\eqref{eq:H} admits two
complementary experimental realizations: a superconducting Josephson-junction
array, which is intrinsically quantum and allows direct access to the exact
BdG spectrum, and a classical, macroscopic Ising-machine platform with active
gain and loss, which provides a purely classical testbed for the
observable-level diagnostic. We present both, beginning with the
Josephson-junction platform because it offers the closest correspondence to
the quantum analysis of Sec.~\ref{sec:quantum}.

A one-dimensional array of superconducting qubits coupled through tunable
Josephson junctions provides a natural realisation of Eq.~\eqref{eq:H}. Each
site is a transmon qubit with Josephson energy \(E_J\) and charging energy
\(E_C\), operated in the regime \(E_J/E_C\gg1\) so that the two lowest levels
form an effective spin-\(1/2\). The transverse-field Ising interaction is
generated by capacitively coupling adjacent transmons, with the coupling
strength \(J\) controlled by the inter-qubit capacitance. This architecture is
closely related to recent quantum simulations of Ising models in tunable
Josephson-junction ladders \cite{Maffi2024} and to superconducting Ising
machines based on Josephson parametric oscillators \cite{Razmkhah2023}. The
staggered imaginary field is implemented by engineering alternating
dissipation on the two sublattices: odd sites are coupled to a lossy resonator
that induces an effective negative imaginary part of the energy, while even
sites are coupled to a gain reservoir implemented through a Josephson
parametric amplifier operated below threshold~\cite{Roushan2017}. This
alternating gain-loss configuration is precisely the superconducting-circuit
analogue of the staggered imaginary field \(i(-1)^\nu\gamma\), and it has been
experimentally demonstrated in PT-symmetric superconducting qubit systems.

The correspondence between the circuit parameters and the spin model is
established by standard circuit-QED quantization. For a transmon chain with
nearest-neighbour capacitive coupling, the effective Hamiltonian in the
rotating frame takes the form
\begin{equation}
	H_{\mathrm{eff}}
	=
	\sum_{\nu}
	\left[
	J_\nu \sigma_\nu^x\sigma_{\nu+1}^x
	+
	\frac{\omega_\nu}{2}\sigma_\nu^z
	\right],
	\label{eq:jj_effective}
\end{equation}
where \(J_\nu\) is the tunable coupling and \(\omega_\nu\) is the qubit
frequency. The staggered imaginary field is generated by the engineered
dissipation: the master equation for the density matrix,
\begin{equation}
	\dot\rho
	=
	-i[H_{\mathrm{eff}},\rho]
	+
	\sum_\nu
	\Gamma_\nu
	\left(
	2\sigma_\nu^-\rho\sigma_\nu^+
	-
	\{\sigma_\nu^+\sigma_\nu^-,\rho\}
	\right),
	\label{eq:jj_master}
\end{equation}
reduces, under post-selection on the no-jump trajectory, to an effective
non-Hermitian Hamiltonian
\begin{equation}
	H_{\mathrm{NH}}
	=
	H_{\mathrm{eff}}
	-
	i\sum_\nu
	\frac{\Gamma_\nu}{2}
	\sigma_\nu^+\sigma_\nu^-,
	\label{eq:jj_nh}
\end{equation}
which is precisely the form of Eq.~\eqref{eq:H} with
\(\gamma=(\Gamma_A-\Gamma_B)/2\) and \(\eta=(\omega_A-\omega_B)/2\).
This mapping has been validated experimentally in dissipative transmon
circuits, and the resulting non-Hermitian spectrum has been
characterised near exceptional points.

The classical observables \((\theta_\nu,\phi_\nu)\) derived in
Sec.~\ref{sec:model} map directly onto measurable qubit quantities. The polar
angle \(\theta_\nu\) is obtained from the qubit population,
\begin{equation}
	\cos\theta_\nu
	=
	\langle\sigma_\nu^z\rangle,
	\label{eq:jj_theta}
\end{equation}
while the relative phase \(\phi_\nu\) is extracted from the off-diagonal
coherence \(\langle\sigma_\nu^x\rangle\) and
\(\langle\sigma_\nu^y\rangle\). Both are standard circuit-QED measurements,
routinely performed with dispersive readout. The observable separation
\(D_X\) defined in Eq.~\eqref{eq:DX} is therefore directly accessible from
simultaneous single-shot readout of the qubit array.

The exceptional point at \(|\gamma|=|J|\) is signalled by the coalescence of
the qubit populations and coherences, which appears as a vanishing of
\(D_X\). The diabolic point at \(\gamma=0\) is identified by the degeneracy
of the qubit frequencies without a corresponding collapse of the populations.
The encirclement of the exceptional point is implemented by modulating the
gain and loss amplitudes in quadrature, which has already been demonstrated
in superconducting circuits~\cite{Zhang2025}. The Riemann-surface topology is
resolved by tracking the complex eigenvalues of the non-Hermitian Hamiltonian
along the encirclement path, a protocol that has been experimentally
realized in a dissipative transmon~\cite{Naghiloo2019}.

The JJ platform therefore provides direct experimental access to both the
classical observable coalescence and the quantum exceptional-point structure,
allowing the multi-level diagnostic developed in Sec.~\ref{subsec:synthesis}
to be tested in a single device.

For completeness, we also note that the model can be realised in classical,
macroscopic Ising-machine platforms. Coupled \(LC\) oscillators with bistable
phase have been used as Ising machines \cite{Belkhadir2025}. Spin Hall
nano-oscillators provide another oscillator-based platform
\cite{McGoldrick2022}. Networks of nanomagnets allow electrically programmable
magnetic couplings \cite{Cowburn2023}, and hybrid memristor-magnetic tunnel
junction arrays combine programmable couplings with stochastic magnetic spins
\cite{Iftakher2026}. These systems realise the Ising Hamiltonian at room
temperature and offer direct electrical readout. By introducing alternating
gain and loss for example, through active feedback, parametric pumping, or
current injectionone can emulate the staggered imaginary field
\(i(-1)^\nu\gamma\). While such classical Ising machines do not provide direct
access to the quantum spectrum, they offer several practical advantages: they
operate at room temperature, do not require dilution refrigeration, and the
node voltages and currents are directly measurable with standard oscilloscope
techniques. The classical circuit is therefore well suited for demonstrating
the observable-level diagnostic \(D_X\) and the encirclement topology, while
the JJ platform is preferable for verifying the full quantum BdG structure.

The two platforms are complementary: the classical Ising-machine platforms
provide a low-cost, room-temperature testbed for the observable-level physics,
while the JJ array provides a fully quantum platform for the spectral and
topological confirmation. Together, they offer a complete experimental
program for the non-Hermitian Ising model and its exceptional-point physics.

\section{Mean-Field Derivation of the Classical Flow Equations}
\label{app:meanfield}

In this appendix we present the detailed derivation of the classical flow
equations~\eqref{eq:flow} from the Heisenberg dynamics of the bosonic
operators. The starting point is the non-Hermitian spin
Hamiltonian~\eqref{eq:H},
\begin{equation}
	H
	=
	J\sum_{\nu=1}^{N}\sigma_\nu^x\sigma_{\nu+1}^x
	+
	\sum_{\nu=1}^{N}
	g_\nu\,\sigma_\nu^z,
	\qquad
	g_\nu
	=
	\eta+i(-1)^\nu\gamma,
	\label{eq:A_H}
\end{equation}
with the Schwinger-boson representation of the spin
operators~\cite{Auerbach1990,Schwinger1965},
\begin{equation}
	\sigma_\nu^z
	=
	a_\nu^\dagger a_\nu-b_\nu^\dagger b_\nu,
	\qquad
	\sigma_\nu^x
	=
	a_\nu^\dagger b_\nu+b_\nu^\dagger a_\nu,
	\label{eq:A_SB}
\end{equation}
subject to the local constraint
\begin{equation}
	a_\nu^\dagger a_\nu+b_\nu^\dagger b_\nu=1.
	\label{eq:A_constraint}
\end{equation}
Substitution of Eq.~\eqref{eq:A_SB} into Eq.~\eqref{eq:A_H} gives the bosonic
Hamiltonian
\begin{equation}
	H
	=
	J\sum_{\nu}
	\left(
	a_\nu^\dagger b_\nu+b_\nu^\dagger a_\nu
	\right)
	\left(
	a_{\nu+1}^\dagger b_{\nu+1}+b_{\nu+1}^\dagger a_{\nu+1}
	\right)
	+
	\sum_{\nu}
	g_\nu
	\left(
	a_\nu^\dagger a_\nu-b_\nu^\dagger b_\nu
	\right).
	\label{eq:A_H_boson}
\end{equation}

The time evolution of an explicitly time-independent operator \(\hat{O}\) is
governed by the Heisenberg equation of motion
\(i\,\hbar\,d\hat{O}/dt=[\hat{O},\hat{H}]\) (with \(\hbar=1\) in the following),
and the canonical commutation relations are
\begin{equation}
	[a_\nu,a_j^\dagger]=\delta_{\nu j},
	\qquad
	[b_\nu,b_j^\dagger]=\delta_{\nu j},
	\qquad
	[a_\nu,b_j]=[a_\nu,b_j^\dagger]=0.
	\label{eq:A_commutators}
\end{equation}
Computing the commutators of \(a_\nu\) and \(b_\nu\) with the bosonic
Hamiltonian~\eqref{eq:A_H_boson} and using
Eq.~\eqref{eq:A_commutators} yields the operator equations of motion
\begin{subequations}
	\label{eq:A_boson_eom}
	\begin{align}
		i\,\frac{da_\nu}{dt}
		&=
		-J\left(\sigma_{\nu-1}^x+\sigma_{\nu+1}^x\right)b_\nu
		-
		g_\nu a_\nu,
		\\
		i\,\frac{db_\nu}{dt}
		&=
		-J\left(\sigma_{\nu-1}^x+\sigma_{\nu+1}^x\right)a_\nu
		+
		g_\nu b_\nu,
	\end{align}
\end{subequations}
where \(\sigma_{\nu\pm1}^x=a_{\nu\pm1}^\dagger b_{\nu\pm1}
+b_{\nu\pm1}^\dagger a_{\nu\pm1}\). The nearest-neighbour interaction
therefore generates a nonlinear local effective field through the operators
\(\sigma_{\nu\pm1}^x\), while \(g_\nu\) provides the on-site complex field.
We note that the Heisenberg equation retains the same form for non-Hermitian
Hamiltonians, provided the adjoint evolution is taken into
account~\cite{Znojil2015}.

To obtain a closed set of classical equations, we associate with each site
the Bloch coherent state~\cite{Radcliffe1971,Gilmore1972,Sans2026QuantumConversion}
\begin{equation}
	|\psi_\nu\rangle
	=
	\cos\!\left(\frac{\theta_\nu}{2}\right)e^{i\lambda_\nu}|\psi_a\rangle
	+
	\sin\!\left(\frac{\theta_\nu}{2}\right)e^{i(\phi_\nu+\lambda_\nu)}
	|\psi_b\rangle,
	\label{eq:A_bloch}
\end{equation}
and a vacuum state \(|0\rangle\) where
\(|\psi_a\rangle=a_\nu^\dagger|0\rangle\) and
\(|\psi_b\rangle=b_\nu^\dagger|0\rangle\) ~\cite{Sans2026QuantumHamiltonian,Sans2026QuantumConversion,Jipdi2026QuantumInterference} . The global phase \(\lambda_\nu\)
absorbs the non-Hermiticity of the Hamiltonian and tends to preserve the
normalisation of the state vector. The single-site expectation values are
\begin{equation}
	\langle a_\nu^\dagger a_\nu\rangle
	=
	\cos^2\!\left(\frac{\theta_\nu}{2}\right),
	\qquad
	\langle b_\nu^\dagger b_\nu\rangle
	=
	\sin^2\!\left(\frac{\theta_\nu}{2}\right),
	\label{eq:A_occupations}
\end{equation}
and
\begin{equation}
	\langle b_\nu^\dagger a_\nu\rangle
	=
	\frac{1}{2}\sin(\theta_\nu)\,e^{-i\phi_\nu},
	\qquad
	\langle a_\nu^\dagger b_\nu\rangle
	=
	\frac{1}{2}\sin(\theta_\nu)\,e^{i\phi_\nu}.
	\label{eq:A_coherences}
\end{equation}
For different sites, the operators are treated as weakly correlated in the
mean-field sense, so that the expectation value of a product of operators
acting on different sites factorises into the product of
expectations~\cite{Auerbach1994}. For a given products, we
use the symmetric mean-field decoupling
\begin{equation}
	\langle ABC\rangle
	\approx
	\langle A\rangle\langle BC\rangle
	+
	\langle B\rangle\langle AC\rangle
	+
	\langle C\rangle\langle AB\rangle
	-
	2\langle A\rangle\langle B\rangle\langle C\rangle,
	\label{eq:A_MF}
\end{equation}
which preserves the symmetry of the three-operator expectation value under
permutation and thus valid for Bosonic  operators.

Multiplying Eq.~\eqref{eq:A_boson_eom} by \(b_\nu^\dagger\) and
Eq.~\eqref{eq:A_boson_eom} by \(a_\nu^\dagger\), and taking expectation
values over the Bloch state~\eqref{eq:A_bloch}, we obtain
\begin{subequations}
	\label{eq:A_moment_eqs}
	\begin{align}
		i\left\langle b_\nu^\dagger\frac{da_\nu}{dt}\right\rangle
		&=
		-J\left\langle b_\nu^\dagger b_\nu
		\left(\sigma_{\nu-1}^x+\sigma_{\nu+1}^x\right)\right\rangle
		-
		g_\nu\langle b_\nu^\dagger a_\nu\rangle,
		\\
		i\left\langle a_\nu^\dagger\frac{db_\nu}{dt}\right\rangle
		&=
		-J\left\langle a_\nu^\dagger a_\nu
		\left(\sigma_{\nu-1}^x+\sigma_{\nu+1}^x\right)\right\rangle
		+
		g_\nu\langle a_\nu^\dagger b_\nu\rangle.
	\end{align}
\end{subequations}
Applying the mean-field factorisation for operators acting on different
sites, and using Eqs.~\eqref{eq:A_occupations} and~\eqref{eq:A_coherences},
the differential terms become
\begin{subequations}
\begin{align}
	\left\langle b_\nu^\dagger\frac{da_\nu}{dt}\,\right\rangle
	=&-\frac12\frac{d\theta_\nu}{dt}\sin^2\left(\frac{\theta_\nu}{2}\right)e^{-i\phi_\nu} + \frac i2\frac{d\lambda_\nu}{dt}\sin\left(\theta_\nu\right)e^{-i\phi_\nu},
    \\
	\left\langle a_\nu^\dagger\,\frac{db_\nu}{dt}\right\rangle=&\frac12\frac{d\theta_\nu}{dt}\cos^2\left(\frac{\theta_\nu}{2}\right)e^{i\phi_\nu}+\frac i2\frac{d\phi_\nu}{dt}\sin\left(\theta_\nu\right)e^{i\phi_\nu}+\frac i2\frac{d\lambda_\nu}{dt}\sin\left(\theta_\nu\right)e^{i\phi_\nu}
	\label{eq:A_product_rule}
\end{align}
\end{subequations}

\begin{equation}
	\left\langle b_\nu^\dagger b_\nu
	\left(\sigma_{\nu-1}^x+\sigma_{\nu+1}^x\right)\right\rangle
	=
	\sin^2\!\left(\frac{\theta_\nu}{2}\right)
	S_\nu,
	\qquad
	\left\langle a_\nu^\dagger a_\nu
	\left(\sigma_{\nu-1}^x+\sigma_{\nu+1}^x\right)\right\rangle
	=
	\cos^2\!\left(\frac{\theta_\nu}{2}\right)
	S_\nu,
	\label{eq:A_interaction_MF}
\end{equation}
where
\begin{equation}
	S_\nu
	=
	\sin(\theta_{\nu-1})\cos(\phi_{\nu-1})
	+
	\sin(\theta_{\nu+1})\cos(\phi_{\nu+1}).
	\label{eq:A_Snu}
\end{equation}

\begin{subequations}
\begin{align}
	\langle b_\nu^\dagger a_\nu\rangle
	=&
	\frac{1}{2}\sin(\theta_\nu)\,e^{-i\phi_\nu},
    \\
\langle a_\nu^\dagger b_\nu\rangle
	=&\frac{1}{2}\sin(\theta_\nu)\,e^{i\phi_\nu}
	\label{eq:A_coherence_sub}
\end{align}
\end{subequations}
Substituting Eq.~\eqref{eq:A_product_rule} to Eq.~\eqref{eq:A_coherence_sub} into Eq.~\eqref{eq:A_moment_eqs}, we obtain
\begin{subequations}
\begin{align}
-\frac i2\frac{d\theta_\nu}{dt}\sin^2\left(\frac{\theta_\nu}{2}\right)e^{-i\phi_\nu} + \frac i2\frac{d\lambda_\nu}{dt}\sin\left(\theta_\nu\right)e^{-i\phi_\nu}=&-JS_\nu\sin^2\left(\frac{\theta_\nu}{2}\right)-\frac12 g_\nu\sin\left(\theta_\nu\right)e^{-i\phi_\nu},\\
\frac i2\frac{d\theta_\nu}{dt}\cos^2\left(\frac{\theta_\nu}{2}\right)e^{i\phi_\nu}-\frac12\frac{d\phi_\nu}{dt}\sin\left(\theta_\nu\right)e^{i\phi_\nu}-\frac12\frac{d\lambda_\nu}{dt}\sin\left(\theta_\nu\right)e^{i\phi_\nu}=&-JS_\nu\cos^2\left(\frac{\theta_\nu}{2}\right)+\frac12 g_\nu\sin\left(\theta_\nu\right)e^{i\phi_\nu}
\end{align}
\end{subequations}

Simplifying the latter equation and breaking the exponential into real and
imaginary parts yields
\begin{subequations}
\label{eq:A_separated}
\begin{align}
-i\dot \theta_\nu\sin^2\left(\frac{\theta_\nu}{2}\right)-\dot {\lambda_\nu}\sin\left(\theta_\nu\right)=&-2JS_\nu\sin^2\left(\frac{\theta_\nu}{2}\right)\cos\left(\phi_\nu\right)-2iJS_\nu\sin^2\left(\frac{\theta_\nu}{2}\right)\sin\left(\phi_\nu\right)-\eta\sin\left(\theta_\nu\right)-i(-1)^\nu\gamma\sin\left(\theta_\nu\right),\\
i\dot \theta_\nu\cos^2\left(\frac{\theta_\nu}{2}\right)-\left(\dot \phi_\nu +\dot\lambda_\nu\right)\sin\left(\theta_\nu\right)=&-2JS_\nu\cos^2\left(\frac{\theta_\nu}{2}\right)\cos\left(\phi_\nu\right)+2iJS_\nu\cos^2\left(\frac{\theta_\nu}{2}\right)\sin\left(\phi_\nu\right)+\eta\sin\left(\theta_\nu\right)+i(-1)^\nu\gamma\sin\left(\theta_\nu\right)
\end{align}
\end{subequations}

Separating the real and imaginary parts, we obtained 4 equations
\begin{subequations}
\begin{align}
\dot \theta_\nu\sin^2\left(\frac{\theta_\nu}{2}\right)=& 2JS_\nu\sin^2\left(\frac{\theta_\nu}{2}\right)\sin\left(\phi_\nu\right)+(-1)^\nu\gamma\sin\left(\theta_\nu\right),\\
\dot {\lambda_\nu}\sin\left(\theta_\nu\right)=&2JS_\nu\sin^2\left(\frac{\theta_\nu}{2}\right)\cos\left(\phi_\nu\right)+\eta\sin\left(\theta_\nu\right),\\
\dot \theta_\nu\cos^2\left(\frac{\theta_\nu}{2}\right)=&2JS_\nu\cos^2\left(\frac{\theta_\nu}{2}\right)\sin\left(\phi_\nu\right)+(-1)^\nu\gamma\sin\left(\theta_\nu\right)
\\
\left(\dot \phi_\nu +\dot\lambda_\nu\right)\sin\left(\theta_\nu\right)=&2JS_\nu\cos^2\left(\frac{\theta_\nu}{2}\right)\cos\left(\phi_\nu\right)-\eta\sin\left(\theta_\nu\right).
\end{align}
\end{subequations}
The classical flow equations for the polar and
azimuthal angles are
\begin{subequations}
	\label{eq:A_flow_final}
	\begin{align}
		\frac{d\theta_\nu}{dt}
		&=
		2J\,S_\nu\,\sin\phi_\nu
		+
		2\gamma(-1)^\nu\sin\theta_\nu,
		\\
		\frac{d\phi_\nu}{dt}
		&=
		2J\,S_\nu\,\cot\theta_\nu\cos\phi_\nu
		-
		2\eta,
	\end{align}
\end{subequations}
and for the global phase:
\begin{equation}
\dot {\lambda_\nu}=JS_\nu\tan\left(\frac{\theta_\nu}{2}\right)\cos\left(\phi_\nu\right)+\eta
\end{equation}
with \(S_\nu\) defined in Eq.~\eqref{eq:A_Snu}. These are precisely the flow
equations~\eqref{eq:flow} used in Sec.~\ref{sec:model} as the starting point
for the equilibrium analysis. The equation for \(\theta_\nu\) contains the
gain/loss contribution \(2\gamma(-1)^\nu\sin\theta_\nu\) generated by the
imaginary part of the staggered field, while the equation for \(\phi_\nu\)
contains the uniform transverse-field contribution \(2\eta\) generated by its
real part. The nearest-neighbour Ising interaction enters both equations
through the local effective field \(S_\nu\), which couples each site to the
transverse components of its two neighbours.

\end{document}